\documentclass[prd,aps,twocolumn,floatfix,nofootinbib]{revtex4}
\usepackage{amsmath}
\usepackage{amssymb}
\usepackage{bbm}
\usepackage{subfigure}
\usepackage{epsfig}
\usepackage{yfonts}

\usepackage{graphicx}   % need for figures

\usepackage{color}      % use if color is used in text
\usepackage{subfigure}  % use for side-by-side figures
\usepackage{hyperref}   % use for hypertext links, including those to external documents and URLs
\usepackage{setspace}

\usepackage{amsthm}
\newcommand{\labell}[1]{\label{#1}}

\newcommand{\be}{\begin{equation}}
\newcommand{\ee}{\end{equation}}
\newcommand{\bea}{\begin{eqnarray}}
\newcommand{\eea}{\end{eqnarray}}
\newcommand{\ba}{\begin{eqnarray}}
\newcommand{\ea}{\end{eqnarray}}
\newcommand{\eg}{{\it e.g.,}\ }
\newcommand{\ie}{{\it i.e.,}\ }

\newcommand{\vv}{\mathbf{v}}
\newcommand{\uu}{\mathbf{u}}

\def\ov{\over}

\begin{document}

\title{Random Matrix Application to Correlations Among Volatility of Assets}
%\smallskip\smallskip\smallskip

\author{Ajay Singh}
\affiliation{
 Perimeter Institute for Theoretical Physics, Waterloo, Ontario N2L 2Y5, Canada
}
\email[]{asingh@perimeterinstitute.ca}

\author{Dinghai Xu}

\affiliation{
Department of Economics, University of Waterloo, Waterloo, Ontario N2L 3G1, Canada
}

%\maketitle
%\vskip .5cm

\begin{abstract}
In this paper, we apply tools from the random matrix theory (RMT) to estimates of correlations across volatility of various assets in the S\&P 500. The volatility inputs are estimated by modeling price fluctuations as GARCH(1,1) process. The corresponding correlation matrix is constructed. It is found that the distribution of a significant number of eigenvalues of the volatility correlation matrix matches with the analytical result from the RMT. Furthermore, the empirical estimates of short and long-range correlations among eigenvalues, which are within the RMT bounds, match with the analytical results for Gaussian Orthogonal ensemble (GOE) of the RMT. To understand the information content of the largest eigenvectors, we estimate the contribution of GICS industry groups in each eigenvector. In comparison with eigenvectors of correlation matrix for price fluctuations, only few of the largest eigenvectors of volatility correlation matrix are dominated by a single industry group. We also study correlations among `volatility return' and get similar results.
\end{abstract}

\maketitle

 \email{ {\rm E-mail:}\ \ {\tt rmyers,$\,$ asingh@perimeterinstitute.ca}}

%\tableofcontents

%%%%%%%%%%%%%%%%%%%%%%%%%%%%%%%
\section{Introduction}

Volatility of asset returns is one of the most important elements in financial research. Since the birth of seminal models like Black-Scholes and the Autoregressive Conditional Heteroscedasticity (ARCH/GARCH), increasing attention has been paid to the analysis of the time-varying behavior in volatilities in the past few decades. Unlike the asset prices, the volatility is latent on the market. In other words, it is not directly observed and therefore, some estimations are necessary to make volatility time series ``visible" for analysis. There are several well known measures to construct volatility. Broadly, we can classify these methods into the following three categories. The first one is the classical parametric model-based methods, referred as the estimated volatility. In particular, the volatility is generated using the models such as ARCH/GARCH, stochastic volatility models, etc. The second type is called the implied volatility, which is backed up by the option pricing formula, \eg Black-Scholes. In the last category, the volatility is constructed non-parametrically from the high frequency trading data, namely realized volatility. All these three volatility measures are extensively used in the financial industry.

In this paper, we want to extend the traditional volatility analysis into a multivariate environment. The volatility correlations are naturally introduced in this picture. Note that the correlations among stock price fluctuations for different assets are very important because of their direct use for risk management in the Markowitz portfolio theory \cite{markowitz, bouchaud}. However, in practice, there are different sources of noise embedded in the final product of estimated correlations, such as finite-sample bias due to the limiting time domain, estimation errors from the inefficiency in the estimating procedure, measurement errors in the model construction process, etc. In their seminal work \cite{bouchaud1}, Laloux {\it et. al.} show that this accumulated noise in the correlation matrix for price fluctuations can be accounted by using the tools from the random matrix theory (RMT) \cite{mehta, randm1, randm2}. In particular, they find that distribution of eigenvalues of empirical correlation matrix, excluding some of the largest eigenvalues, fits very well in the Mar\u{c}enko-Pastur distribution of the RMT \cite{mehta, randm1, randm2, sengupta}. In \cite{stanley1, stanley2}, it is further shown that the properties of this correlation matrix resembles with the Gaussian Orthogonal ensemble (GOE). These results strongly suggest that eigenvalues of correlation matrix falling under the Mar\u{c}enko-Pastur distribution contain no genuine information about the financial markets. Hence, one should systematically filter out such noise from the correlations for a more accurate estimation of future portfolio risk (see \cite{bouchaud3} and references therein).

The correlations among volatility of different assets are useful in portfolio selection, pricing option book and in certain multivariate econometric models to forecast prices and volatility \cite{implied, multivariate}. For example in the context of the Black-Scholes model, the variance of a portfolio $\pi$ of options exposed only to vega risk  is given by \cite{implied}
\be
\textrm{Var}(\pi) \,=\, \sum\limits_{i,j,k,l} w_i w_l \Lambda_{ij} \Lambda_{lk} \mathbb{C}_{jk}  \,.
\labell{implied}
\ee
Here ${w}_i$ are weights in the portfolio, $\mathbb{C}_{ij}$ is the correlation matrix for implied volatility of underlying assets and the vega matrix $\Lambda_{ij}$ is defined as
\be
\Lambda_{ij}\,=\,{\partial p_i \ov \partial \nu_j}\,,
\ee
where $p_i$ is the price of option $i$ and $\nu_j$ is the implied volatility of asset underlying option $j$. In volatility arbitrage strategies, generally correlations among the `volatility return', that is the change in volatility, are used. Since volatility correlation matrix $\mathbb{C}_{ij}$ in \eqref{implied} or the correlations among volatility return are always estimated, they contain both systematic and random errors. Hence, for a better forecast of risk, one certainly needs to estimate and remove noise from these correlation matrices.

The use of volatility correlation matrix in risk management and volatility arbitrage strategies can have another important aspect. Once volatility correlation matrix is obtained, one can ask several interesting questions about its eigenvalues and eigenvectors. In both risk management and arbitrage strategies for assets or derivatives, one tries to utilize as much information as available about the market. For example, eigenvectors of correlation matrix for price fluctuations reveal that the most correlated structures in the stock market, which are also stable for a longer time period, are the industry sectors \cite{stanley2, stanley5, stanley4}. In this case, by selecting a portfolio vector orthogonal to all the relevant eigenvectors, one can significantly reduce the portfolio risk. Now in case of volatility correlation matrix, one can naturally ask if its eigenvectors carry any new information about the market and whether they are stable in time. If yes, then how can this be utilized for a better estimation of risk and improving volatility arbitrage strategies?

In this paper, we apply tools from the RMT to volatility correlation matrix. We use one of the well studied econometric models, GARCH$(1,1)$, to estimate the time evolution of volatility of assets \cite{arch, garch, tsay}. In the GARCH$(1,1)$ process, the volatility is measured as the standard deviation of price fluctuations. In econometrics literature, we do realize that there are rather sophisticated models available to measure daily volatility. However, it has been observed that if one is not concerned about the asymmetric response of volatility to price fluctuations, \ie leverage effect \cite{leverage}, GARCH(1,1) is not outperformed by any other model  to a significant level \cite{garch11, andersen}. Hence, GARCH (1,1) is treated at least as a starting point of the analysis. We leave the use of multivariate models and use of other proxies of volatility for future work.

The rest of the paper is organized as follows. In section \ref{simulation}, we discuss the data being used in this study and how do we model the price fluctuations to generate volatility time series. Using volatility time series, we construct the correlation matrix and calculate its eigenvalues in section \ref{rmt}. Then, by fitting the cumulative probability distribution of eigenvalues with analytical expression, we find the optimum number of eigenvalues which fall under the Mar\u{c}enko-Pastur distribution of RMT. Once we know which eigenvalues could possibly be pure noises, we perform further tests to ensure that they are indeed random. We study statistical properties of these eigenvalues in section \ref{eigenvalues}. The nearest-neighbor spacing distribution, next to nearest-neighbor spacing distribution and number variance for unfolded eigenvalues are calculated. While first two quantities test the short-range correlations among the eigenvalues, the later evaluates the long-range correlations. We find a very good agreement of these quantities with analytical results for GOE of the RMT. In section \ref{eigenvectors}, we study the eigenvector statistics and find that the eigenvector corresponding to the largest eigenvalue is the `market mode'. Since the eigenvalue for market mode is of the order of the total number of assets, market has very strong correlations across volatility of assets. As we discuss in section \ref{evecdis}, the market mode also influences the eigenvectors under the Mar\u{c}enko-Pastur distribution. We further calculate the contribution of GICS industry groups in the eigenvectors which are supposed to carry genuine information. The eigenvectors of correlation matrix for price fluctuations, which are outside the Mar\u{c}enko-Pastur distribution, are relatively stable in time and are dominated by particular industry groups \cite{stanley5,stanley4,stanley3}. However, in case of volatility correlation matrix, only few of the largest eigenvectors are dominated by particular industries. In section \ref{robust}, we discuss the robustness of our results by modeling the time series with small and larger number of parameters compared to GARCH(1,1). Finally in section \ref{dis}, we summarize our results and discuss future directions. In appendix \ref{volreturn}, we mention results for application of the RMT to correlation matrix for volatility return \footnote{It is worth mentioning that in this paper three types of correlation matrices are discussed: return correlation matrix defined as \eqref{eq1x6}, volatility correlation matrix given by \eqref{eq1x4} and correlation matrix for volatility returns in \eqref{cx2}.}.

%%%%%%%%%%%%%%%%%%%%%
\section{Estimates of volatility and correlation matrix} \labell{simulation}

In this paper, we use daily closing prices for 427 stocks in the S\&P 500 over the time period covering from July 1, 2009 to June 28, 2013 \cite{yahoo}. We represent the total number of stocks by $N$ ($N=427$) and the length of the time series for price fluctuations by $T$ ($T=1005$). The companies omitted are those that left or joined the S\&P 500 list in this time duration \footnote{For few time series, we observe that either they are non-stationary or the GARCH(1,1) is not a good model to estimate the volatility. Hence, these time series are also not considered in the paper.} Note that our data belongs to a time duration which contains the aftershocks of economic crisis of year 2008-09. It has been observed that market is generally strongly correlated in volatile periods \cite{bouchaud4}. Hence, more eigenvalues of return and volatility correlation matrices tend to deviate from the standard RMT results, as compared to analysis in \cite{bouchaud1, stanley2}.

To model each time series, we begin with defining the return on stock $i$ at time $t$ by $r_{i,t} = \log(P_{i,t+1}/P_{i,t})$, where $P_{i,t}$ is the price of stock $i \in\{1,2,\dots,N\}$ at time $t\in\{1,2,\dots,T\}$. We can normalize the return time series such that they have unit variance and zero mean, and write as an $N\times T$ matrix $G_{it}$ such that
\be
G_{it} \, =\, { r_{i,t} - \bar{r}_{ i} \ov \sigma_i } \,,
\labell{eq1x70}
\ee
where
\be
\bar{ r}_i \,=\, \langle r_{i,t} \rangle \quad \textrm{and } \quad \sigma_i \,=\, \sqrt{\langle (r_{i,t} - \bar{r}_i )^2 \rangle}\,.
\labell{eq1x7}
\ee
In our notation, angle brackets represent the average over the time series until unless stated explicitly. Now the correlation matrix for price fluctuations can be written as
\be
C \,=\, {1\ov T} GG^{T} \,.
\labell{eq1x6}
\ee
For convenience, we refer $C$ as return correlation matrix.

Further, we estimate volatility time series by modeling the return $r_{i,t}$ using a univariate GARCH(1,1) process \footnote{It is worth mentioning that unit root tests are preformed to verify the stationarity condition before fitting the data into the GARCH structure \cite{forecast, autoarima1}.}. Generally in a GARCH framework, both the conditional mean $\bar{r}_{i,t}$ and conditional variance ${\sigma}_{i,t}^2$ at $t$, given the information $I_t$, are functions of time:
\bea
\bar{r}_{i,t} &\,=\,& \langle r_{i,t}|I_t \rangle_e \,, \\
 {\sigma}_{i,t}^2 &\,=\,& \langle (r_{i,t}-\bar{r}_{i,t}^2 | I_t \rangle_e \,.
\labell{eq1x2}
\eea
Here $\langle \rangle_e$ represents the average over the ensemble and $I_t$ is information about the prices till time $t$.

%Now, we test the return time series for heteroscedasticity by calculating autocorrelation of $(r^i_t)^2$ and find that it does not cutoff abruptly. Since this is one of the crucial properties that GARCH models are expected to capture, the time series $r_t^i$ is modeled as GARCH(1,1) process \cite{tsay, garch11, andersen}:

In this paper, a standard GARCH(1,1) structure is used. There are two equations in the process to model the conditional mean and conditional volatility,
\bea
r_{i,t} &\,=\,& {\sigma}_{i,t} \, \epsilon_{i,t}  \,, \notag \\
{\sigma}_{i,t}^2  &\,=\,& \alpha^i_0 + \alpha^i_1 \, r_{i,{t-1}}^2 + \beta^i_1 \, {\sigma}_{i,{t-1}}^2 \,,
\labell{eq1x3}
\eea
where $\epsilon_{i,t}$ is a random element drawn from a Gaussian or $t$-distribution. The coefficients $\alpha^i_0$, $\alpha^i_1$ and $\beta^i_1$ for stock $i$ are estimated using standard econometric packages \cite{rugarch}. Using these parameters, we can sequentially estimate the volatility time series $\sigma_{i,t}$ based on \eqref{eq1x3}.

\begin{figure}
\centering
\includegraphics[width=.5\textwidth]{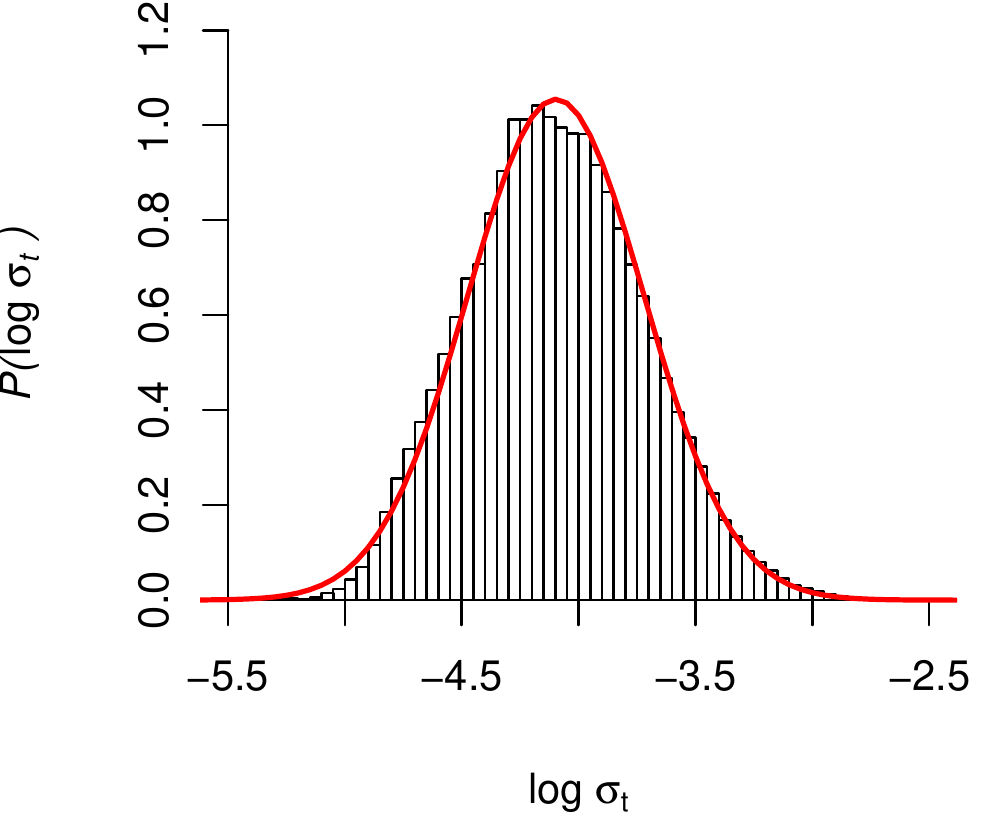}
\caption{(Colour Online) The probability density distribution for logarithm of daily volatility, \ie $\log({{\sigma}}_{i,t})$ is shown. The volatility time series are generated by modeling the price fluctuations for 427 stocks from S\&P 500 as univariate GARCH(1,1) processes. The data set has skewness $0.19$ and kurtosis $3.01$. The red line is the Gaussian fit with mean $-4.097\pm .001 $ and standard deviation $0.378\pm.001$.}
\label{fig1a}
\end{figure}

As it is generally observed, we find that volatility has a distribution very close to a lognormal. In figure \ref{fig1a}, we draw the distribution of $\log(\sigma_{i,t})$, which fits well in Gaussian distribution with mean $-4.099\pm .001 $ and standard deviation $0.379\pm.001$. We can again normalize the volatility time series such that it has zero mean and unit variance:
\be
\widehat{\sigma}_{i,t} \,=\,  \frac{{\sigma}_{i,t} - \bar{\sigma}_i }{s_i}\,.
\labell{eq1x8}
\ee
Here $\bar{\sigma}_i$ and $s_i$ are mean and standard deviation of the volatility time series ${\sigma}_{i,t}$:
\be
\bar{\sigma}_i \,=\, \langle \sigma_{i,t} \rangle \quad \textrm{and} \quad s_i\,=\, \sqrt{\langle ({\sigma}_{i,t} - \bar{\sigma}_i)^2 \rangle} \,.
\labell{eq1x5}
\ee
The positive tail in the distribution of $\widehat{\sigma}_{i,t}$ fits well with a power-law coefficient 4.5, which is close to 5.4 for daily mean absolute deviation of high-frequency return at an interval of 5 minutes \cite{bouchaud}. Now, we can arrange the time series for normalized volatility $\widehat{\sigma}_{i,t}$ as $N\times T$ matrix $\mathbb{G}$ and then the volatility correlation matrix is given by
\be
\mathbb{C} \,=\, \frac{1}{T} \mathbb{G} \mathbb{G}^T \,.
\labell{eq1x4}
\ee
Using the volatility correlation matrix, one can compute its eigenvalues $\lambda_i$ and conjugate eigenvectors $\vv_i$. Now in the next section, we find the optimum number of eigenvalues that fit in the density distribution from the RMT.

%%%%%%%%%%%%%%%%%%%%%
\section{RMT and Eigenvalue distribution}
\labell{rmt}

In this section, we compare the eigenvalue distribution for volatility correlation matrix \eqref{eq1x4} with the analytical results from the RMT. Note that correlation matrix has $N(N-1)/2$ independent components and one requires time series of length $T>N$ for estimation of correlation matrix. However, if time series are uncorrelated, an empirical estimation of correlations require time series of infinite length. In the context of the RMT, assume that we have $N$ uncorrelated time series of length $T$, with random elements drawn from a Gaussian distribution with zero mean and standard deviation $s_0$. We can arrange these time series in an $N\times T$ matrix $R$. For these time series, we can calculate correlation matrix, that is a Wishart matrix $RR^T/T$, and distribution of its eigenvalues. In the limit $N \to \infty$ and $T\to \infty$, such that $Q=T/N$ is fixed, the distribution of eigenvalues becomes the well-known Mar\u{c}enko-Pastur distribution \cite{marcenko, mehta}:
\be
P^{\textrm{\tiny RM}}(\lambda) \,=\, {Q \ov 2\pi s_0^2} { \sqrt{(\lambda_{+}-\lambda)(\lambda - \lambda_{-})}  \ov \lambda}\,,
\labell{eq2x1}
\ee
where
\be
\lambda_{\pm} \,=\, s_0^2 \left(1+{1\ov Q} \pm {2 \ov \sqrt{Q}} \right) \,.
\labell{eq2x2}
\ee
where $\lambda$ represents eigenvalues and $\lambda_- \leq \lambda \leq \lambda_+$. Both the values $\lambda_\pm$ are greater than zero and only when $Q\to1$, the gap between zero and $\lambda_-$ disappears and we recover Wigner semi-circle law. For finite $T$ and $N$, the abrupt cut-off at both ends $\lambda_\pm$ are replaced by rapidly decaying tails. Note that \eqref{eq2x1} is exact when the elements of random, uncorrelated time series have Gaussian distribution. If random elements are drawn from a power-law distribution outside the L\'{e}vy stable range, the eigenvalue distribution is still found to be in good agreement with \eqref{eq2x1} \cite{stanley2, randm1}. Since volatility time series roughly follows a lognormal distribution, we have explicitly constructed $N$ time series of length $T$ with random elements drawn from a lognormal distribution. We find that \eqref{eq2x1} is consistent with the eigenvalue distribution of correlation matrix for these artificial time series.

Further, as the correlations are introduced among the time series, some eigenvalues begin to move out of the bulk distribution \eqref{eq2x1} \cite{sengupta}. These eigenvalues carry genuine information about the market. In this case, effective standard deviation $s_0$ in \eqref{eq2x1} differs from its original value for the time series. We can refer the bulk of the distribution by `noise' as it represents no-information states. However, the eigenvalues outside this regime correspond to genuine correlations and the associated eigenvectors represent the correlated segments of the market.

In the next section, we show how certain eigenvalues of the correlation matrix form the bulk of the distribution and these fit well with the analytical expression \eqref{eq2x1} from the RMT. It has been observed that distributions similar to \eqref{eq2x1} can arise even when time series have well-defined correlations. Consequently, we perform further checks on the eigenvalues falling under the Mar\u{c}enko-Pastur distribution in section \ref{eigenvalues} and compare the results with analytical expressions for GOE.

%%%%%%%%%%%%%%%%%%%%
\subsection{Eigenvalue density} \labell{eval}

\begin{figure}
\centering
\includegraphics[width=.5\textwidth]{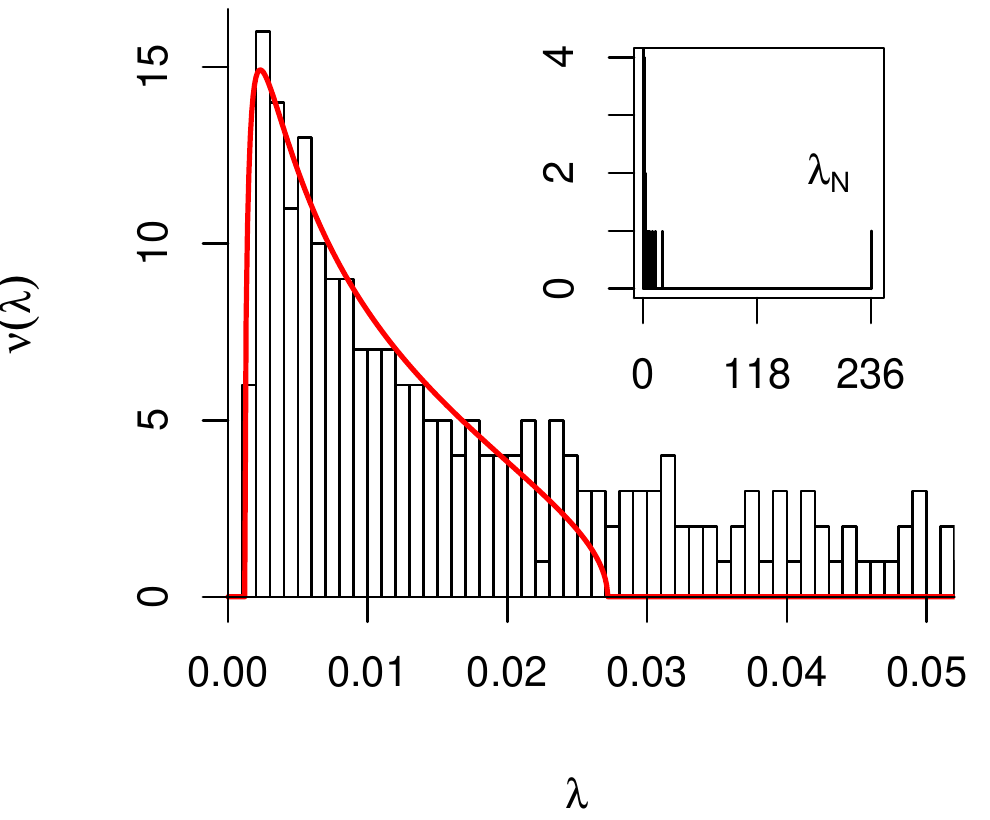}
\caption{(Colour Online) We draw histogram for eigenvalues of volatility correlation matrix. The y-axis is the frequency of eigenvalues $\nu(\lambda)$ in bins of width .001. The smooth plot is the analytical fit \eqref{eq2x1} from the RMT. This is being estimated by fitting the data points \eqref{eq2x3} in the cumulative probability distribution \eqref{eq2x3x1}. The parameters of the fit are $Q=2.351$, $s_0^2=0.09952 \pm 0.00005$ and $\alpha=0.3890 \pm 0.0014$. We have further multiplied the estimated plot with a factor $(.001\sum_{\lambda<\lambda_+} \nu(\lambda)/\alpha)$ to match the frequency scale on y-axis. (Inset) We show that the largest eigenvalue $\lambda_N=237$ is much larger than the bulk of the distribution.}
\label{fig2x1}
\end{figure}

\begin{figure}
\centering
\includegraphics[width=.5\textwidth]{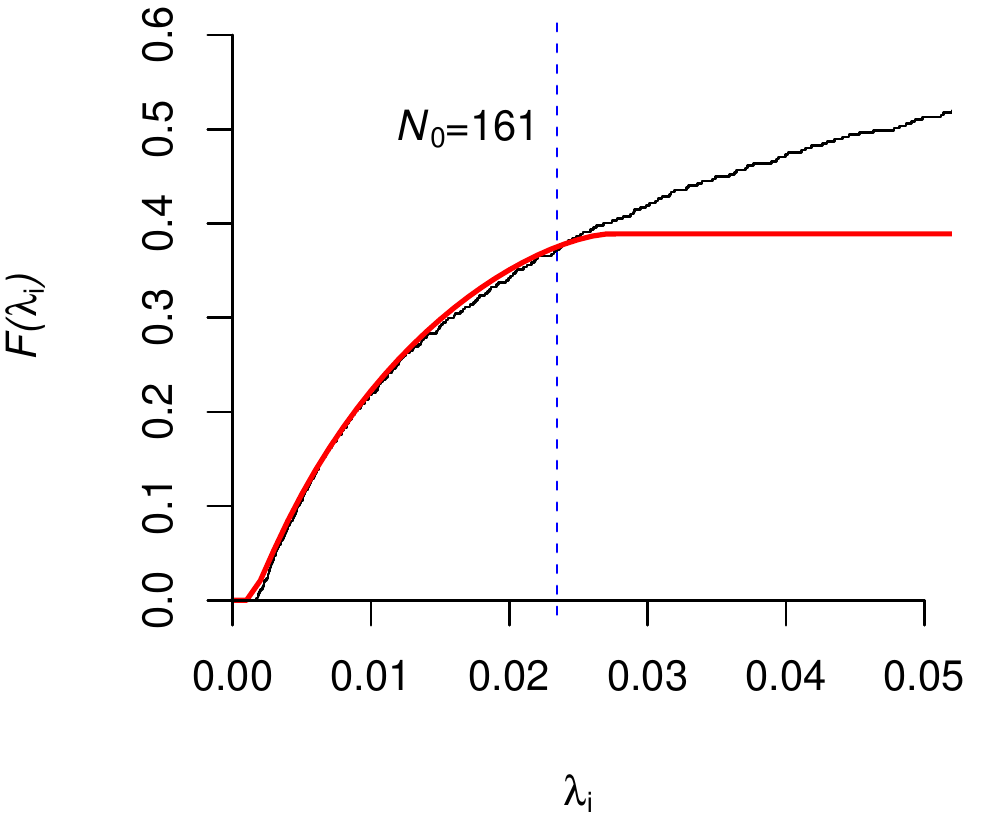}
\caption{(Colour Online) The staircase function is the empirical cumulative probability distribution \eqref{eq2x3} for eigenvalues of volatility correlation matrix. The smooth plot is the analytical fit \eqref{eq2x3x1} for $Q=2.351$, $s_0^2=0.09952 \pm 0.00005$ and $\alpha=0.3890 \pm 0.0014$. To generate this fit, first $N_1=161$ eigenvalues are being used and this edge is shown with a vertical dashed line.}
\label{fig2a}
\end{figure}

Using volatility correlation matrix \eqref{eq1x4}, we can calculate its eigenvalues and sort them in a sequence as $\{\lambda_1, \lambda_2, \dots, \lambda_N\}$, where $\lambda_1$ is the smallest and $\lambda_N$ is the largest eigenvalue. The histogram of eigenvalues is shown in figure \ref{fig2x1}. First we notice that the smallest eigenvalue $\lambda_1=.0015$, which is positive. Second, the tail of the distribution is decaying smoothly, instead of ending abruptly. We also notice that the largest eigenvalue $\lambda_N$ is approximately 237, much larger than the values in the bulk of the distribution. These observations are pretty consistent with the characteristics in our sample data set in which the market was found to be quite volatile and strongly correlated. Note that since the trace of the correlation matrix is fixed to be $N$, when $\lambda_N$ becomes larger, the peak of the distribution moves closer to zero.

To fit the bulk of the eigenvalue distribution in \eqref{eq2x1}, we use the empirical cumulative distribution function,
\be
F(\lambda_i) \,=\, {1\ov N} \sum \limits_{j=1}^N \Theta(\lambda_i-\lambda_j) \,.
\labell{eq2x3}
\ee
Here $\Theta(\lambda)$ is the Heaviside step function and we have plotted $F(\lambda_i)$ as a function of eigenvalues in figure \ref{fig2a}. Since only a subset of eigenvalues are noise, we fit part of $F(\lambda_i)$ in
\be
F^{\textrm{\tiny RM}}(\lambda) \,=\, \alpha \int_{-\infty}^\lambda d\lambda' \, P^{\textrm{\tiny RM}}(\lambda', s_0) \,,
\labell{eq2x3x1}
\ee
with parameters $s_0$ and $\alpha$, keeping $Q=T/N$ fixed. As shown in figure \ref{fig2x1}, a significant number of eigenvalues are outside the bulk of the distribution. Hence, $\alpha$ is introduced to take care of the normalization in \eqref{eq2x3x1} when it is compared with only part of empirical cumulative distribution $F(\lambda_i)$ in \eqref{eq2x3}. In addition, when several eigenvalues are outside the RMT fit, the effective standard deviation $s_0$ changes. In particular if time series are weakly correlated, there are only a few eigenvalues greater than $\lambda_+$, that is the theoretical bound \eqref{eq2x2} from the RMT. Then, the effective standard deviation is approximately
\be
s_0^2 \approx 1 - {1\ov N}\sum \limits_{i=N_0+1}^{N} \lambda_i  \,,
\labell{eq2x4}
\ee
where $N_0$ is the number of eigenvalues which fall under the Mar\u{c}enko-Pastur distribution \eqref{eq2x1}. 

\begin{figure}
\centering
\includegraphics[width=.5\textwidth]{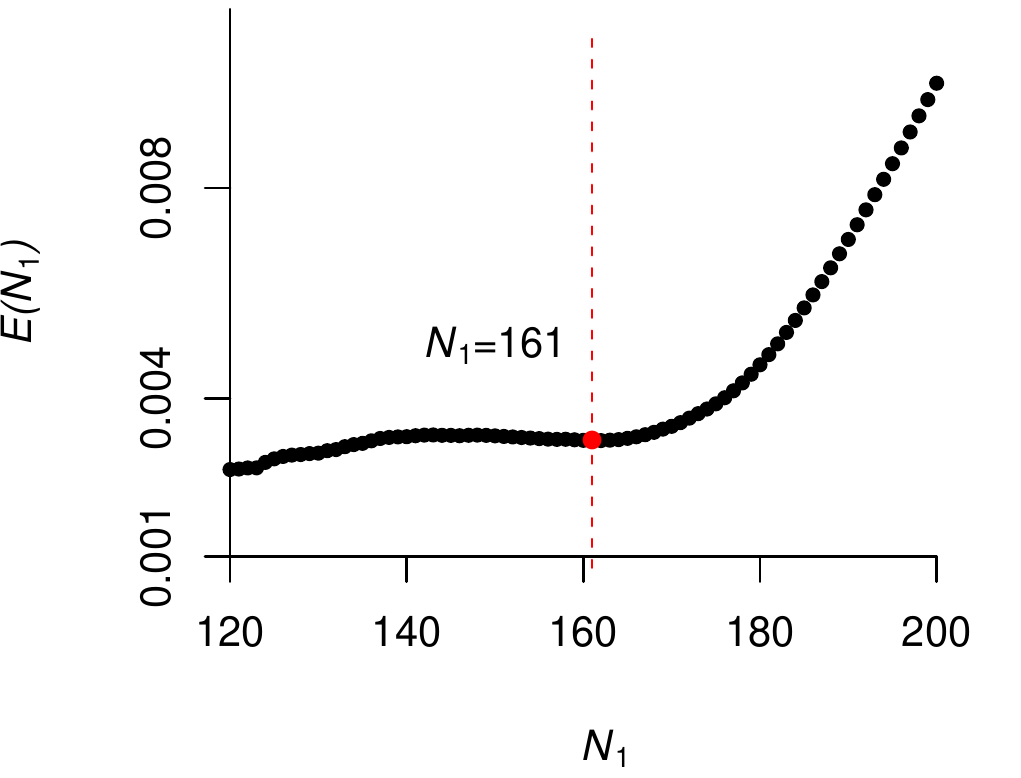}
\caption{(Colour Online) We plot the root-mean-square error (RMSE) of the best RMT fit in $N_1$ data points of empirical cumulative probability distribution \eqref{eq2x3}. As we change $N_1$, we approach a local minimum at $N_1=161$. This minimum is pointed out by a vertical dashed line.}
\label{fig2b}
\end{figure}

To fit the data in \eqref{eq2x3x1}, we first choose a subset of $N_1$ eigenvalues $\{\lambda_1,\lambda_2,\dots,\lambda_{N_1}\}$ and corresponding data points in empirical cumulative distribution \eqref{eq2x3}. Then we fit these data points in \eqref{eq2x3x1} and estimate $\alpha$ and $s_0$ by minimizing the root-mean-square error (RMSE). This minimized RMSE for selected $N_1$ data points can be represented by,
\be
E(N_1) \,=\, \underset{\alpha,s_0}{\operatorname{min}} \left\{ \sqrt{{ {1\ov N_1} \sum\limits_{i=1}^{N_1}\big[F(\lambda_i) - F^{\textrm{\tiny RM}}(\lambda_i,\alpha,s_0)\big]^2 }} \right\}\,.
\ee
We draw $E(N_1)$ as a function of $N_1$ in figure \ref{fig2b}. It appears that there is a local minimum in $E(N_1)$ around $N_1=161$ and beyond this threshold, $E(N_1)$ begins to increase almost linearly. In this way, we find an `optimum' number of eigenvalues which are used to fit in the RMT result \eqref{eq2x3x1}. Note that the value of $N_1$ is sensitive to the fitting procedure and the indicator being used. However, the precise distribution of noise have a decaying tail, instead of a sharp edge as in \eqref{eq2x1}. Hence a minor change in $N_1$ do not overestimate noise.

For $N_1=161$, the estimated parameters are $s_0^2=0.009952 \pm 0.00005$ and $\alpha=0.3890 \pm 0.0014$. Using \eqref{eq2x2}, we also estimate $\lambda_-=.0012$ and $\lambda_+=.0272$. For these values of parameters, the probability density \eqref{eq2x1} is shown in figure \ref{fig2x1}. Finally, we find that there are $N_0=173$ eigenvalues such that $\lambda_i \leq \lambda_+$ and they fall under the Mar\u{c}enko-Pastur distribution. We notice that the estimated values of $s_0^2$ and $\alpha$ are close to what we get from \eqref{eq2x4}, that is $s_0^2 \approx 0.00429$ and $\alpha\approx N_0/N=.4051$.

%%%%%%%%%%%%%%%%%%%%%%%%
\section{Statistical properties of eigenvalues} \labell{eigenvalues}

In previous section, we investigated which eigenvalues fall under the Mar\u{c}enko-Pastur distribution and possibly are pure noise. However, it is quite possible for time series to contain genuine correlations and then have a distribution very similar to the Mar\u{c}enko-Pastur distribution. For this reason, we perform further diagnostic tests to ensure that the eigenvalues bounded by the RMT limits are indeed noise. Since correlation matrix is a real symmetric matrix, we use the data to evaluate the null hypothesis that it belongs to GOE of the RMT. To do so, we compare the short and long-range correlations among eigenvalues with analytical results for GOE. 

%Another motivation to consider GOE is that it has been shown that properties of return correlation matrix \eqref{eq1x6} resemble with GOE \cite{stanley1}. Since volatility is the second moment of return, it is more likely that the volatility correlation matrix belongs to GOE than Gaussian unitary ensemble (GUE) or Gaussian symplectic ensemble (GSE).

By definition, GOE of random matrices has two important properties. First, if $M$ is a real symmetric matrix and an element of GOE, all of its elements are statistically independent. Second, the ensemble is invariant under the orthogonal transformation. In other words, any transformation of an element $M \to \mathcal{O}^T M \mathcal{O} $, where $\mathcal{O}$ is a real orthogonal matrix, leaves the joint probability of elements of $M$ invariant. Now because of this symmetry, the elements of GOE display some universal properties. Since, some of these properties are self-averaging, one can observe these by studying a single, large element of GOE.
%So we consider the null hypothesis that volatility correlation matrix is an element of GOE, and compare the statistical properties of eigenvalues and eigenvectors with universal properties of GOE.
In particular, we study the short-range correlations by calculating the nearest-neighbor and next to nearest-neighbor spacing distributions for unfolded eigenvalues. We also compare long-range correlations among eigenvalues by calculating the number variance. Even for a small number of eigenvalues falling within the RMT bounds, that is $N_0=172$, we find a very good agreement with universal properties of GOE.

%%%%%%%%%%%%%%%%%%%%%%%
\subsection{Nearest-neighbor spacing distribution} \labell{nnbr}

\begin{figure}
\centering
\includegraphics[width=.5\textwidth]{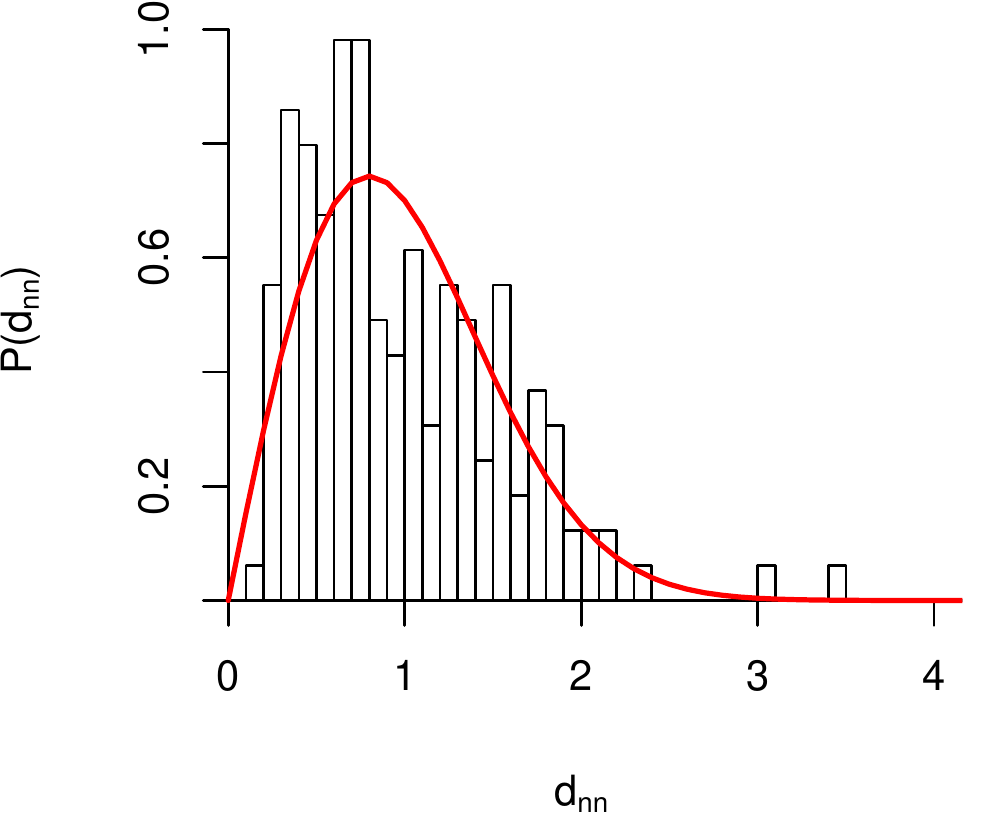}
\caption{(Colour Online) We draw histogram for nearest-neighbor spacing distribution of Gaussian unfolded eigenvalues. The smooth plot is the fit $\beta_1P_{\textrm{\tiny GOE}}$ with $\beta_1=0.98 \pm0.05$ and $P_{\textrm{\tiny GOE}}$ is given by \eqref{eq2x5}. The Kolmogorov-Smrinov statistics and corresponding p-value with reference to analytical expression \eqref{eq2x5} are consecutively 0.066 and 0.48.}
\label{fig3a}
\end{figure}

The first test for GOE is the distribution of nearest-neighbor spacing for unfolded eigenvalues of volatility correlation matrix. To unfold the eigenvalues, we use the technique of Gaussian broadening as it is used in the context of Hubbard model in \cite{broadening}. The Gaussian unfolding procedure is briefly summarized in appendix \ref{appa}. We consider all the $N_0$ eigenvalues within the theoretical bounds \eqref{eq2x2} and calculate the unfolded eigenvalues $\xi_i$. By definition \eqref{ax3}, the unfolded eigenvalue is a map from $\lambda_i$ to $\xi_i$ such that, $\xi_i$ has a uniform distribution. Now we estimate the distribution for nearest-neighbor spacing $d=(\xi_{i+1}-\xi_i)$ and it is shown in figure \ref{fig3a}. One of the standard results from the RMT is that the distribution of nearest-neighbor spacing of unfolded eigenvalues for GOE is the famous Wigner surmise \cite{mehta, randm1, randm2}:
\be
P_{\textrm{\tiny GOE}}(d) \,=\, {\pi d \ov 2} \exp\left( -{\pi \ov 4}d^2 \right) \,.
\labell{eq2x5}
\ee
We fit the estimated density of nearest-neighbor spacing in $(\beta_1 P_{\textrm{\tiny GOE}})$ with normalization parameter $\beta_1$. As shown in figure \ref{fig3a}, the empirical data fits well in analytical expression for GOE. We find $\beta_1= 0.98 \pm 0.05$, which is very close to the exact value $\beta_1=1$. We further find that for nearest-neighbor spacing distribution, the Kolmogorov-Smirnov statistics is 0.066 and p-value is 0.48. At the significance level of .05, p-values for Kolmogorov-Smirnov test with reference to their distributions discard GUE and GSE.

%%%%%%%%%%%%%%%%%%%%%%%
\subsection{Next to nearest-neighbor spacing distribution} \labell{nnnbr}

\begin{figure}
\centering
\includegraphics[width=.5\textwidth]{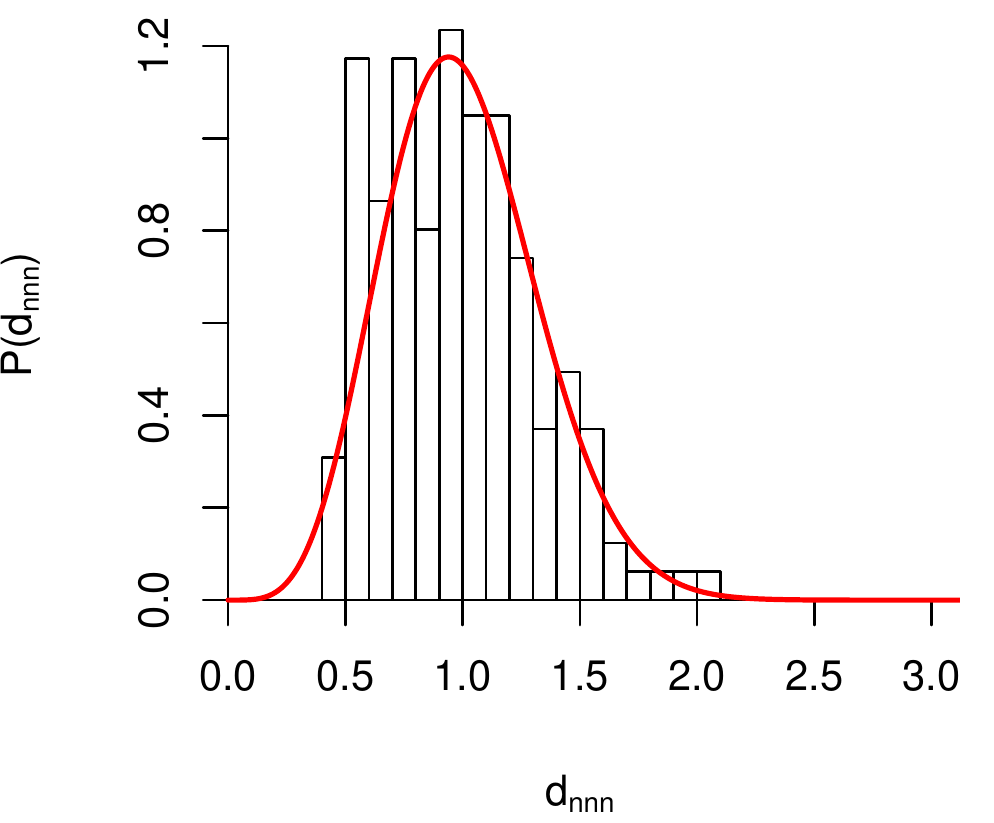}
\caption{(Colour Online) We draw histogram for next to nearest-neighbor spacing distribution of Gaussian unfolded eigenvalues. The smooth plot is the fit $\beta_2P_{\textrm{\tiny GSE}}$, where $\beta_2=0.96 \pm 0.05$ and $P_{\textrm{\tiny GSE}}$ is given by \eqref{eq2x6}. The Kolmogorov-Smrinov statistics and corresponding p-value with reference to analytical distribution \eqref{eq2x6} are 0.063 and 0.55.}
\label{fig3b}
\end{figure}

The second test of GOE is to compare the next to nearest-neighbor spacing distribution of unfolded eigenvalues with RMT results. For GOE, the distribution of next to nearest-neighbor spacing of unfolded eigenvalues is shown to be equivalent to nearest-neighbor spacing distribution for GSE \cite{mehta, randm1, randm2}. This is called the GSE test of GOE. The analytical expression for nearest-neighbor spacing for GSE is,
\be
P_{\textrm{\tiny GSE}}(d) \,=\, {2^{18} \ov 3^6\pi^3} d^4 \exp\left( -{64 \ov 9 \pi} d^2 \right) \,.
\labell{eq2x6}
\ee
To calculate empirical values of next to nearest-neighbor spacing, we select all the eigenvalues within RMT bounds \eqref{eq2x2}. We further divide these eigenvalues $\lambda_i$ in two sets with even and odd index $i$. Now, both sets are such that next to nearest-neighbor eigenvalues of original sequence are in the same groups. For each set, following the procedure in appendix \ref{appa}, we perform Gaussian broadening and calculate unfolded eigenvalues $\xi^\textrm{\tiny even/odd}$. Using these unfolded eigenvalues, we calculate the nearest-neighbor spacings $d^\textrm{\tiny even/odd}=(\xi_{i+1}^\textrm{\tiny even/odd} - \xi_i^\textrm{\tiny even/odd})$ in each set. Now we combine data from both of the sets to get probability density for next to nearest-neighbor spacing in original sequence of eigenvalues. The density distribution is shown in figure \ref{fig3b} and we fit it in $(\beta_2 P_{\textrm{\tiny GSE}})$ with the normalization constant $\beta_2$. We find that $\beta_2=0.96 \pm 0.05$, very close to the exact value one. The Kolmogorov-Smirnov statistics for next to nearest-neighbor distribution is 0.063 and p-value can not discard the null-hypothesis at the significance level of 0.5.

%%%%%%%%%%%%%%
\subsection{Number variance} \labell{number}

In this section, we compare long-range correlations among eigenvalues with the analytical results for GOE. There are examples of systems, for which the Hamiltonion do not belong to GOE but short-range correlations in the spectrum do resemble with short-range correlations in GOE \cite{randm1}. Hence, to establish the fact that the eigenvalues under the Mar\u{c}enko-Pastur distribution are pure noise, it is required to compare the long-range correlations among eigenvalues with analytical results for GOE. One of the quantities that probes long-range, two-level correlations among the eigenvalues is the number variance. It is defined as the variance of number of unfolded eigenvalues in the interval of length $\ell$ around each unfolded eigenvalue $\xi_i$ \cite{mehta, randm1, randm2}:
\be
\Sigma(\ell)^2 \,=\, {1 \ov N_0}\sum_{i=1}^{N_0} (n(\xi_i,\ell) - \langle n(\xi_i,\ell) \rangle_\xi )^2  \,.
\labell{eq2x7}
\ee
Here $N_0$ is number of unfolded eigenvalues and $n(\xi_i, \ell)$ is the number of eigenvalues in the interval $[\xi_i - \ell/2, \xi_i + \ell/2]$. Further, $\langle n(\xi_i,\ell) \rangle_\xi$ is the average of number of eigenvalues in the interval $[\xi_i - \ell/2, \xi_i + \ell/2]$ and here averaging is done over the unfolded eigenvalues $\xi_i$. Since unfolded eigenvalues have a uniform distribution, $\langle n(\xi_i,\ell) \rangle_\xi=\ell$. Now if the spectrum is translation invariant, the number variance can be written as
\be
\Sigma(l)^2 \,=\, \ell - 2 \int_0^\ell dr\, (\ell-r) Y_2(r) \,,
\ee
where $Y_2(r)$ is related to the two-point correlations. If there are no long-range correlations among eigenvalues, one gets the Poisson spectrum with $Y_2=0$ and $\Sigma(\ell)^2=\ell$. However, the expression of $Y_2$ for GOE takes the following form
\be
Y_2(r) \,=\, y(r)^2 + {dy(r) \ov dr} \int_r^\infty dr' y(r') \,,
\ee
where
\be
y(r) \,=\, {\sin(\pi r)  \ov \pi r} \,.
\ee

\begin{figure}
\centering
\includegraphics[width=.5\textwidth]{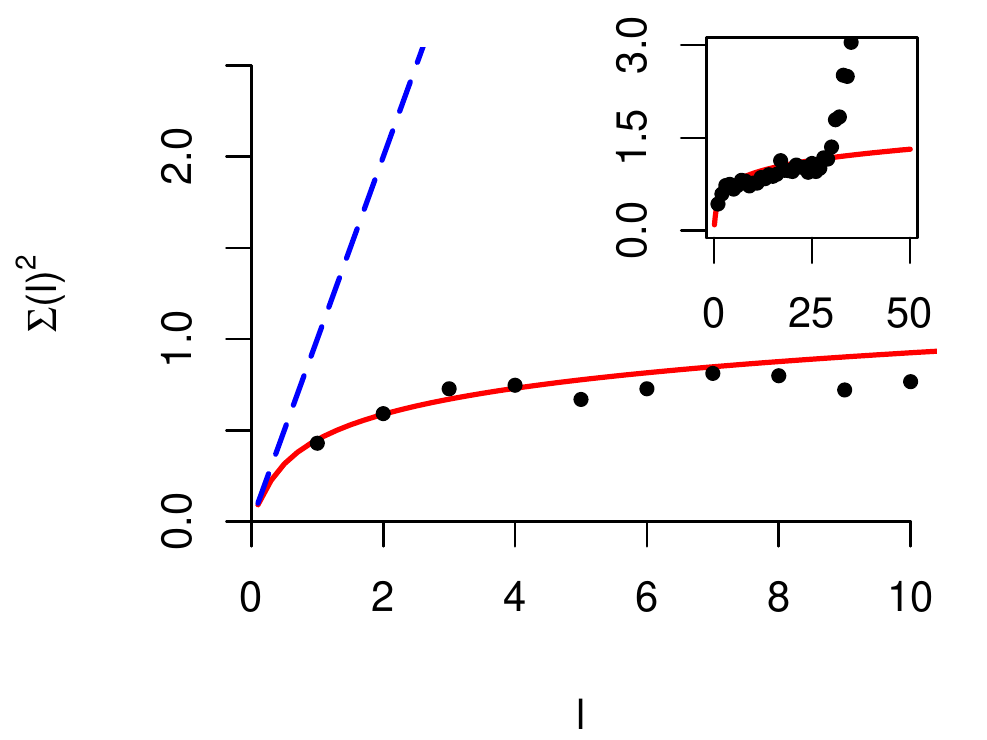}
\caption{(Colour Online) We plot number variance as a function of spacing parameter $\ell$. The black points are the estimated number variance for the eigenvalues bounded by the theoretical edge $\lambda_+$.  The smooth line is the theoretical value of number variance for GOE. The dashed line is number variance for the uncorrelated Poisson eigenvalues, that is $\Sigma^2=\ell$. (Inset) Black points are estimated number variance and smooth line is theoretical value of number variance for GOE. We can see that for large values of $\ell$, the estimated number variance diverge from GOE. This is generally observed when we are dealing with a finite number of eigenvalues.}
\label{fig4}
\end{figure}

To estimate number variance empirically, we unfold the eigenvalues using the Gaussian broadening procedure in appendix \ref{appa}. Since number variance takes into account the long-range correlations, it is affected by both of the edges at $\lambda_-$ and $\lambda_+$, particularly when $N_0$ is not very large compared to $\ell$. Hence, we estimate number variance only using the eigenvalues deep in the bulk of the distribution. The empirical estimates and theoretical value of number variance \eqref{eq2x7} for GOE are shown in figure \ref{fig4}. We find that for $\ell \leq 10$, the fit is in very good agreement with the exact result. However, for larger values of $\ell$, as it is shown in the inset, number variance diverges to Poisson spectrum. This behavior is common for the cases where the number of eigenvalues is not very large.

The results from the sections \ref{nnbr}, \ref{nnnbr} and \ref{number} show that the short and long-range correlations for eigenvalues of volatility correlation matrix resemble very well with the analytical results for GOE of the RMT. These results strongly support the hypothesis that eigenvalues of the correlation matrix, which fall under the Mar\u{c}enko-Pastur distribution, are pure noise. In the next section, we investigate the properties of eigenvectors of volatility correlation matrix and compare these with the eigenvectors of return correlation matrix.

%%%%%%%%%%%%%%
\section{Eigenvector statistics} \labell{eigenvectors}

In this section, we study the properties of eigenvectors of volatility correlation matrix. We first recall some results for eigenvectors of return correlation matrix \eqref{eq1x6}. In case of price fluctuations, it has been observed that most of the eigenvalues fall under the Mar\u{c}enko-Pastur distribution and only few of them are outside the bulk. The components of the eigenvectors, which are conjugate to noisy eigenvalues, have Gaussian distribution \cite{bouchaud1, stanley1,stanley2}. The eigenvector conjugate to the largest eigenvalue is the `market mode'. This mode is equivalent to a portfolio in which every asset is equally weighted. Furthermore, other larger eigenvalues, which are outside the theoretical edges from the RMT, are expected to carry genuine correlations. Few of the largest eigenvectors are found to be stable in time over a duration as long as ten years \cite{stanley5, stanley4}. However, as one moves from the largest to some smaller eigenvalues, the time duration of stability reduces and eventually eigenvectors become random. To understand the information contained in these largest eigenvectors, one of the approaches is to decompose them in industry groups \cite{stanley5, stanley4}. It has been observed that while most of the small eigenvectors randomly distribute the weight to all the industries, the largest eigenvectors are dominated by one or two sectors. In section \ref{ecosec}, we follow the same approach to study the properties of the eigenvectors of volatility correlation matrix.

In section \ref{evecdis}, first we discuss the distribution of eigenvectors of volatility correlation matrix. We find that similar to return correlation matrix, the eigenvector conjugate to the largest eigenvalue is the market mode. Then we focus on the distribution of eigenvectors conjugate to the eigenvalues within the RMT bounds. Since the largest eigenvalue $\lambda_N$ of correlation matrix is of the order of the total number of eigenvalues $N$, we find that it significantly affects the other eigenvectors. However, once we remove the effect of the market mode, we find that the eigenvector distribution is Gaussian, consistent with the RMT. Next, we investigate on the information content of the eigenvectors conjugate to the largest eigenvalues in section \ref{ecosec}. After removing the effect of the common market mode, we estimate the weights of different industry groups in these eigenvectors. Interestingly, we find that compared to return correlation matrix, very few of the largest eigenvectors of volatility correlation matrix are dominated by a few industry groups.

%%%%%%
\subsection{Distribution of eigenvectors} \labell{evecdis}

\begin{figure}
\centering
\includegraphics[width=.5\textwidth]{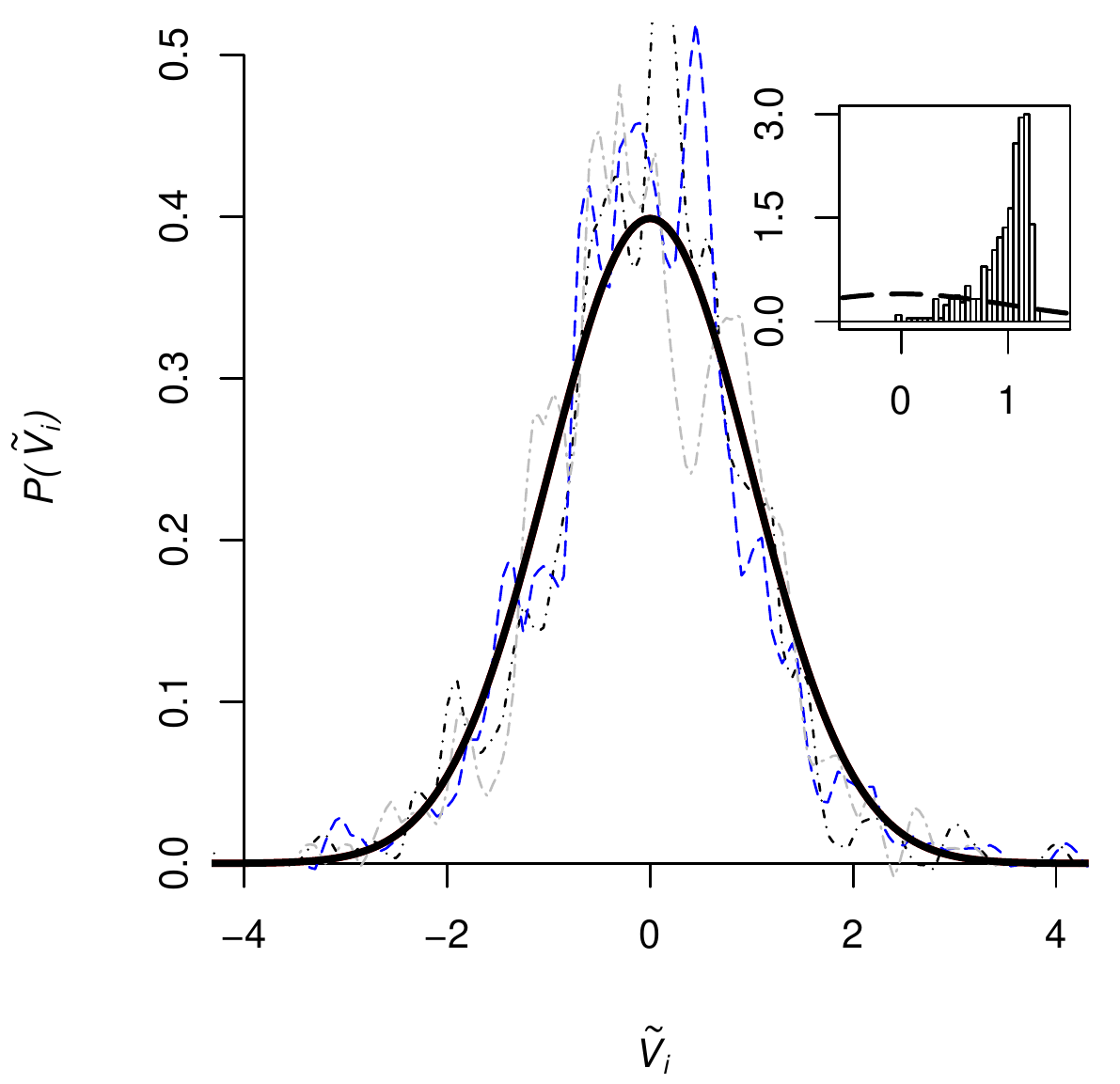}
\caption{(Colour Online) The solid black line is the Gaussian distribution for eigenvectors of random matrices. The dotted and dashed lines are distributions for few eigenvectors $\tilde{\vv}_i$ of volatility correlation matrix, which fall deep into the Mar\u{c}enko-Pastur distribution. These eigenvectors are obtained after removing the effect of the market mode from volatility time series and fit very well in Gaussian distribution. (Inset) We show the distribution of components of the market mode. The dashed line is the Gaussian distribution for a completely random eigenvector. }
\label{fig5}
\end{figure}

To study the eigenvector statistics, we first normalize each eigenvector $\vv_i$, associated with eigenvalue $\lambda_i$, such that $\vv_i^T\vv_i=N$. We find that for the largest eigenvalue $\lambda_N$, most of the components of associated eigenvector are clustered around one. This eigenvector is the market mode and its distribution is shown in the inset of figure \ref{fig5}. Now we examine the eigenvectors conjugate to the eigenvalues falling under the Mar\u{c}enko-Pastur distribution. According to the RMT, these eigenvectors should have a Gaussian distribution. However, we find that the distribution is peaked more sharply than a Gaussian distribution. We also observe similar behavior in the return eigenvectors but it is not as strong. Since the largest eigenvalue in both cases are much bigger in comparison with eigenvalues in the bulk, market mode has a significant influence on other eigenvectors \cite{stanley5,stanley4,marsili2}. Hence, it is reasonable to remove the effect of the market mode and re-examine the eigenvector distribution.

To remove the effect of the market mode, we regress the volatility time series on the market mode variable and use the residual to re-calculate the correlation matrix \cite{stanley5, stanley4}. If the market mode is the eigenvector $\vv_N$, we can write the volatility time series for the market as
\be
M \,=\, \vv_N^T\mathbb{G} \,=\, \sum \limits_{i=1}^N \vv_{N,i} \mathbb{G}_{it} \,.
\labell{5x1}
\ee
$\mathbb{G}$ is an $N\times T$ matrix containing time series for normalized volatility $\widehat{\sigma}_{i,t}$, as defined in \eqref{eq1x8}. To remove the influence of the market mode, which is a common factor to all the assets, we construct the following regression,
\be
\widehat{\sigma}_{i,t} \,=\, \alpha^i + \beta^i M_t +\varepsilon_{i,t}\,,
\labell{5x2}
\ee
where $\alpha^i$ and $\beta^i$ are stock specific constants and the residual $\varepsilon_i$ is such that $\langle \varepsilon_i \rangle =0$ and $\langle \varepsilon_i M\rangle =0$. The estimated residual from the above regression is used to construct the corresponding correlation matrix $\tilde{\mathbb{C}}$, its eigenvalues $\tilde{\lambda}_i$ and eigenvectors $\tilde{\vv}_i$.

There are several observations. First, we find that one of the eigenvalues, which was related to the market mode earlier, is zero. We also observe that after a proper rescaling, the eigenvalues $\tilde{\lambda}_i$ under the Mar\u{c}enko-Pastur distribution are quite close to their original values ${\lambda}_i$. To see this, we recall that the sum of eigenvalues $\tilde{\lambda}_i$ is $N$, \ie $\sum \tilde{\lambda}_i=N$. Previously for original correlation matrix $\mathbb{C}$, the sum of the eigenvalues, excluding the largest eigenvalue $\lambda_N$, was $(N-\lambda_N)$. Now if we homogeneously rescale the new eigenvalues such that their sum is $(N-\lambda_N)$, we find that $\tilde{\lambda}_i(N-\lambda_N)/N$ are reasonably close to $\lambda_i$. One can also see this rescaling from the change in the effective variance of the original time series. We give more details on this in appendix \ref{appb}.

After removing the common market factor, we find that the eigenvector distribution for $\tilde{\vv}_i$ fits reasonably well with a Gaussian distribution. We have shown distribution of several eigenvectors in the figure \ref{fig5}. Note that the eigenvectors are normalized such that $\tilde{\vv}_i^T\tilde{\vv}_i=N$. In this figure, the thick black line is the Gaussian distribution with variance one.

%%%%%%%%%
\subsection{Industry groups and comparison with return eigenvectors} \labell{ecosec}

\begin{figure}
\centering
\includegraphics[width=.5\textwidth]{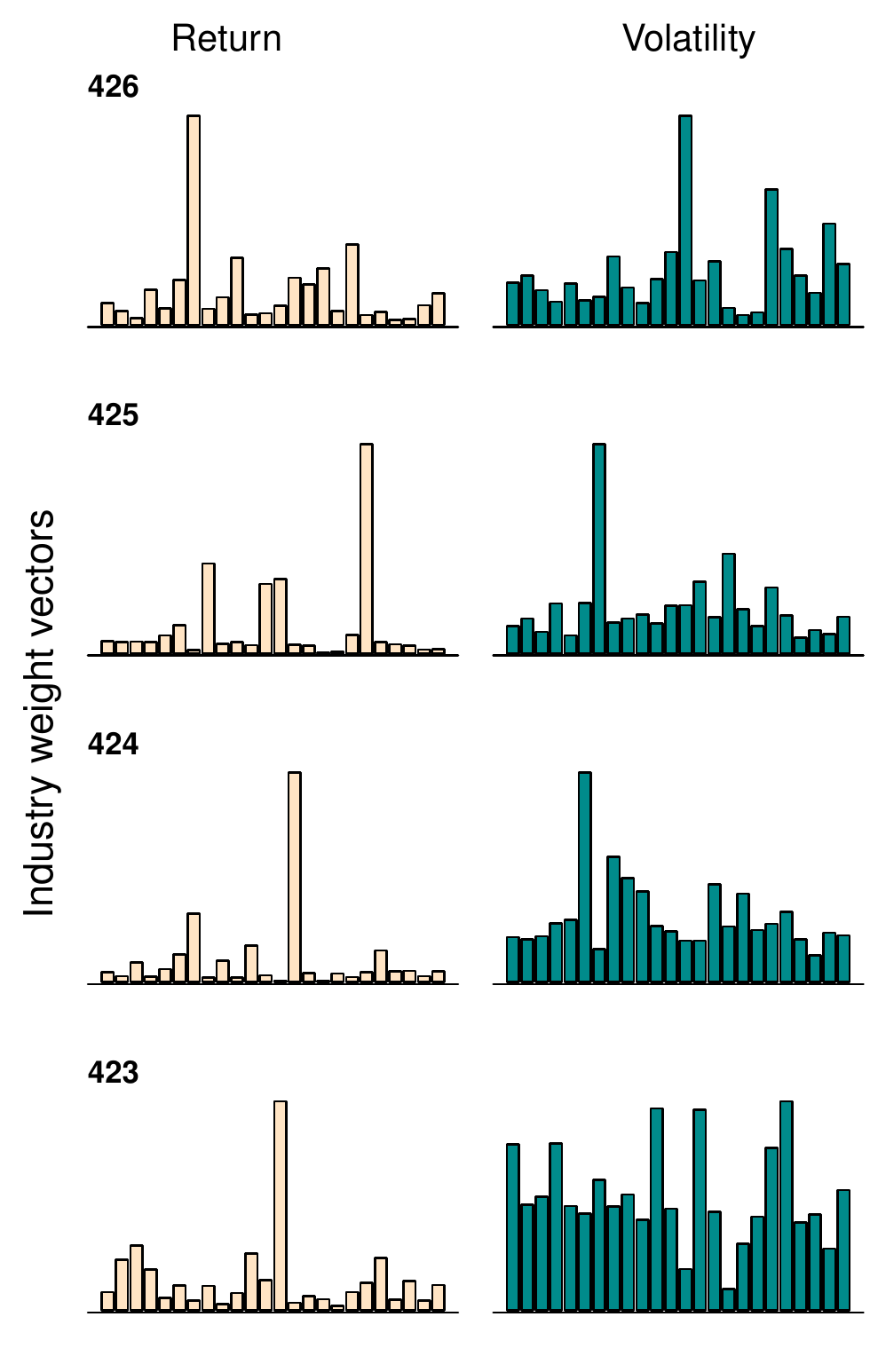}
\caption{(Colour Online) In left column, the components of weight vectors \eqref{5x4} for four largest eigenvalues of return correlation matrix, excluding the market mode, are shown. The left column contains the weight vectors for four largest eigenvalues of volatility correlation matrix. By comparing the eigenvectors 423, we can quickly observe that volatility eigenvector is not dominated by a single industry group, however return eigenvector is. We also point out the industry groups that have the largest contributions in these eigenvectors. For return, the eigenvectors and GICS industry groups are following: 426 -- utilities, 425 -- banks, 424 -- energy, 423 -- real estate. For volatility eigenvectors, the largest industry groups are: 426 -- real estate, 425 -- utilities, 424 -- semiconductors and semiconductors equipments, 423 -- consumer services.}
\label{fig7}
\end{figure}

\begin{figure}
\centering
\subfigure[]{\label{fig6a}\includegraphics[width=.5\textwidth]{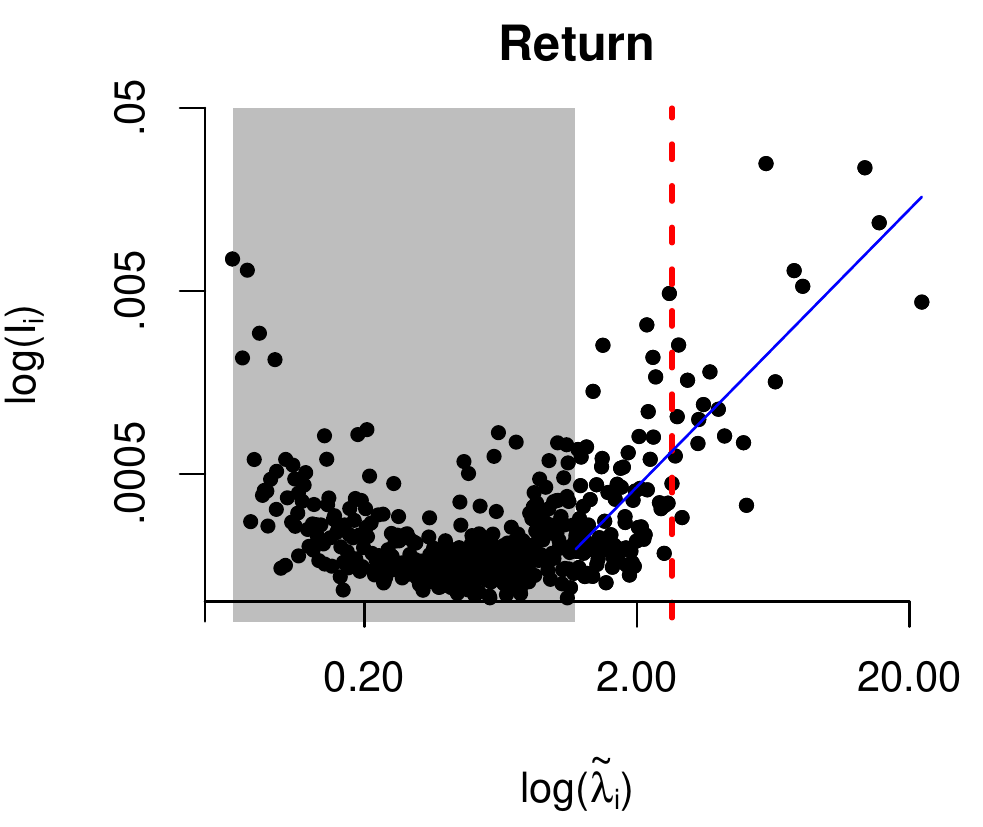}}
\subfigure[]{\label{fig6b}\includegraphics[width=.5\textwidth]{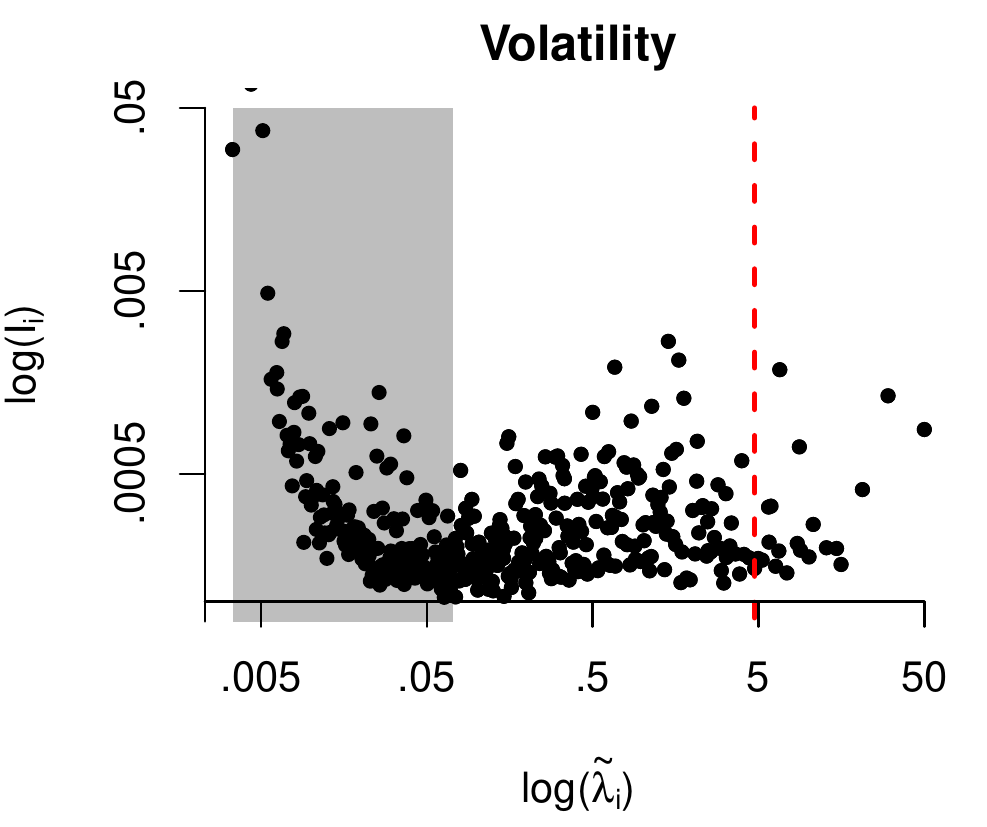}}
\caption{(Colour Online) Panel (a) shows inverse participation ratio of industry weight vectors \eqref{5x4} for return eigenvectors as a function of $\tilde{\lambda}_i$ on a log-log plot. The vertical dashed line shows the position of eigenvalue $\tilde{\lambda}_{407}$ and there are twenty eigenvalues on the right hand side of this. The linear fit in inverse participation ratio for eigenvalues outside the RMT bounds has a slop 1.52, as it is shown by a solid line in the plot. Panel (b) shows inverse participation ratio of weight vectors for volatility eigenvectors as a function of $\tilde{\lambda}_i$ on log-log plot. Again, the vertical line shows the location beyond which, the twenty largest eigenvalues fall. In this case, we can see that compared to return correlation matrix, only few of the largest eigenvalues have large inverse participation ratio and are dominated by a few industry groups.}
\label{fig6}
\end{figure}

In this section, we compare the information contained in the relevant eigenvectors of volatility correlation matrix with that of the return correlation matrix. The main finding is that very few of the volatility eigenvectors, which are supposed to carry genuine information about the market, are dominated by the industry groups as compared to the corresponding return eigenvectors. This result is consistent with the observation that in financial markets, it is much harder to diversify the volatility risk for portfolios \footnote{We thank Samuel Vazquez for pointing this out to us.}. We first estimate the contributions of GICS industry groups in the eigenvectors. Then, by comparing inverse participation ratio of industry contributions, we observe that relatively fewer volatility eigenvectors receive dominating contributions from particular industry groups.

Using the time series \eqref{eq1x70} for normalized returns, we calculate the correlation matrix \eqref{eq1x6}. We follow the procedure described in the section \ref{rmt}. It is found that $75\%$  eigenvalues of return correlation matrix fall under the analytical bounds from the RMT. The movement of the market hides several correlations among its components \cite{stanley5,stanley4,marsili2}. Hence, to see the presence of industry groups in eigenvectors, we remove the influence of the market mode using the technique discussed in \eqref{5x2}. In the cleaned-up volatility and return eigenvectors, we then calculate the contributions of different industry groups as follows.

We classify $427$ companies in the GICS industry groups using a four-digit code system. 24 groups are achieved by the classification. Each group has $n_a$ companies, where $a=1,2,\dots,g$ with $g=24$.  The number of companies in each group, that is $n_a$, range from $4$ to $42$. Now, we can define a $g\times N$ projection matrix $\mathbb{P}$, which estimates the fraction of each industry contributing in the eigenvectors \cite{stanley5, stanley4}. The entries in the projection matrix are,
\be
\mathbb{P}_{ai} \,=\,\left\{
\begin{array} {l l}
1/n_a & \textrm{ if stock $i$ is in group $a$ } \\
0 &  \textrm{ otherwise}
\end{array}
\right. \,.
\labell{5x3}
\ee
In $\mathbb{P}$, row $a$ is a vector with a weight $1/n_a$ to all the $n_a$ companies in group $a$. We also define a vector $\uu_i$ conjugate to each eigenvector $\tilde{\vv}_i$ such that its components are square of the later, \ie $\uu_{i,j}= \tilde{\vv}_{i,j}^2 $. Note that $\tilde{\vv}_i$ represents eigenvectors of correlation matrices after removing the effect of the market mode. The projection matrix acts on vectors $\uu_i$ and gives $g$-dimensional weight vector $\rho_i$:
\be
\rho_i \,=\, \gamma_i \mathbb{P}\, {\uu_i}\,.
\labell{5x4}
\ee
Here $\gamma_i$ is the normalization constant such that $\sum_{a=1}^g \rho_{i,a}=1$. Ideally for the market mode $\vv_N$, all the components of $\rho_N$ should be $1/g$. In figure \ref{fig7}, we compare four weight vectors of return and volatility correlation matrix. Now we can simply use inverse participation ratio $I_i$,
\be
I_i \,=\, \sum \limits_{a=1}^g \rho_{i,a}^4 \,,
\labell{5x5}
\ee
of the weight vectors $\rho_i$ as an indicator of the dominance of a single industry in corresponding eigenvector $\tilde{\vv}_i$. If an eigenvector is dominated by a single industry, then in the weight vector, all the elements will be zero excluding one. In this case inverse participation ratio will be one. However, if all the industry groups contribute equally in an eigenvector, the inverse participation ratio will become $1/g^3$.

In figure \ref{fig6a} and \ref{fig6b}, we plot inverse participation ratio of weight vectors on log-log plot for return and volatility correlation matrices. The x-axis is eigenvalues $\tilde{\lambda}_i$, that we get after removing influence of the market mode from normalized return and volatility time series. The gray area indicates the region bounded by the RMT limits \eqref{eq2x2} in both cases. In \cite{stanley5, stanley4}, authors studied return correlation matrix for 1000 stocks and observed that eigenvectors related to twenty largest eigenvalues were almost stable on the time scale of a year. However, as one moves to the smaller eigenvalues, the eigenvectors become more and more unstable over time. In figure \ref{fig6a} and \ref{fig6b}, we have drawn vertical dashed lines to indicate the position, beyond which the twenty largest eigenvalues fall. %of eigenvalue $\tilde{\lambda}_{N-20}$.

Now, we concentrate on investigating these twenty largest eigenvectors. The structure of industry weights in eigenvectors are apparently different between return and volatility eigenvectors. As shown in figure \ref{fig6a}, in case of return correlation matrix, the largest eigenvalues have large inverse participation ratio and they are dominated by only a few industry groups. In fact outside the RMT bounds, inverse participation ratio appears to follow a power-law as a function of eigenvalues with an exponent 1.52. Deep within the RMT bounds, the inverse participation ratio is of the order of $1/g^3$. This indicates that all the industry groups randomly contribute in these eigenvectors. 

Correspondingly, in the case of volatility correlation matrix in figure \ref{fig6b}, the number of eigenvectors with small inverse participation ratio is quite large as compared to the return correlation matrix. For the sake of comparison, let us set a benchmark value of inverse participation ratio $I_0=1/12^3$. This is the inverse participation ratio for a hypothetical weight vector such that it contains equal contribution from half of the industry groups and no contribution from rest. Sixteen out of twenty largest volatility eigenvectors have inverse participation ratio less than this benchmark value. For return eigenvectors, this number is only two. We also notice that for some smallest eigenvectors of return and volatility correlation matrices, the inverse participation ratio is large. Consistent with \cite{stanley2}, we observe that these eigenvectors are `localized' and receive large contributions from a few stocks in a particular industry group.

The above observations suggest that the eigenvectors of volatility correlation matrix do not carry as much information about the industry groups as the return correlation matrix does. However, it is quite possible that our approach to remove effect of the market mode is inadequate and strong non-linear effects due to the market still hide information about the industry groups in volatility eigenvectors. To confirm that the residual time series $\varepsilon_i$ in \eqref{5x2} are indeed independent of the market mode $M$, we can use a quantity called generalized kurtosis \cite{bouchaud}. The idea is to first transform the residuals $\varepsilon_i$ and the market mode $M$ to $\widehat{\varepsilon}_i=F_i^{-1}(\varepsilon_i)$ and $\widehat{M}=F_M^{-1}(M)$, such that distribution of each $\widehat{\varepsilon}_i$ and $\widehat{M}$ is a Gaussian with unit variance. Now if ${\varepsilon}_i$ and $M$ are independent, then so are $\widehat{\varepsilon}_i$ and $\widehat{M}$. To test this assumption behind our linear model, we can study the generalized kurtosis,
\be
\kappa_i \,=\, \langle \widehat{\varepsilon}_i^2 \, \widehat{M}^2 \rangle - \langle \widehat{\varepsilon}_i^2 \rangle \, \langle \widehat{M}^2 \rangle - 2 \langle \widehat{\varepsilon}_i \, \widehat{M} \rangle \,,
\labell{5x6}
\ee
which should be very small if \eqref{5x2} succeeds if removing effect of the market mode. We can further define the following measure
\be
K \,=\, {1 \ov N} \sum \limits_{i=1}^{N} \kappa_i \,,
\labell{5x7}
\ee
as a merit of any model in this context. The value of this indicator for volatility eigenvectors is $K_v=0.082$ and for return eigenvectors, it is $K_r=0.115$. Both of these values are reasonably small and indicate that model \eqref{5x2} does succeed in removing the effect of the market mode. In fact, $K_v<K_r$ indicates that in comparison to return, this model works better for volatility!

The results in this section strongly suggest that the interpretation of largest eigenvectors of volatility correlation matrix in terms of industry groups is inadequate. In our opinion, this could be attributed to following two reasons. First, it might be that excluding the largest two or three eigenvectors of the volatility correlation matrix, other eigenvectors are quite unstable in time. Another possibility could be that, volatility divides the market in sectors which differ from the industry groups. A better understanding of time evolution and information content of largest eigenvectors of volatility correlation matrix will certainly have important applications in volatility risk management. In this context, it would be interesting to apply tools from cluster analysis \cite{mantegna1, mantegna2, marsili1, coron1} and further study the time evolution of eigenvectors along the line of \cite{stanley5, stanley4, allez1, allez2, fenn, conlon}.

%%%%%%%%%%%%%%%%%%%%%%%%%%
\section{Robustness of results} \labell{robust}

In this section, we do some robust diagnostic analysis on our results. In our procedure, we estimate the volatility time series by modeling 427 individual return time series as GARCH(1,1) processes. In total, there are $3\times 427$ estimated parameters in the pool. We explore two different regimes of number of parameters. First, we consider a single GARCH(1,1) model for all the 427 time-series
\bea
r_t &\,=\,& \sigma_t \epsilon_t  \,, \notag \\
\sigma_t^2  &\,=\,& \alpha_0 + \alpha_1 r_{t-1}^2 + \beta_1 \sigma_{t-1}^2 \,.
\labell{6x1}
\eea
Here $r_t$ is the return at time $t$ and $\epsilon_t$ is a random variable drawn from student's $t$-distribution. Coefficients $\alpha_0$, $\alpha_1$ and $\beta_1$ are parameters to be estimated. The estimation is done using the standard maximum likelihood estimation procedure with the joint log-likelihood function $\mathcal{L}$ defined as follows,
\be
\mathcal{L} \,=\, \sum \limits_{i=1}^N \mathcal{L}_i(\alpha_0, \alpha_1, \beta_1|r_i) \,.
\labell{6x2}
\ee
Here $\mathcal{L}_i(\alpha_0, \alpha_1, \beta_1|r_i)$ represents the log-likelihood function for individual return time series $r_{i,t}$ for stock $i$.

%Now we estimate the values of the parameters $\alpha_0$, $\alpha_1$ and $\beta_1$ such that they maximize this joint log-likelihood function. At first, it might seem strange to fit a single GARCH in all the time series. However, t

The motivation of doing this exercise comes from the observation that for $N=427$, the largest eigenvalue for the return correlation matrix is around 193. This implies that the conjugate eigenvector, \ie the market itself, is strongly correlated. Hence, one can assume that a universal GARCH model governs the market. We do find that all the estimated parameters are statistically significant. Repeating the calculations in sections \ref{rmt}, \ref{eigenvalues} and \ref{eigenvectors}, we find that although there are small changes in the numerical values, the main results are qualitatively the same.

Our second experiment is that instead of fitting return time-series in a GARCH(1,1), as in section \ref{simulation}, we model the each time series as a more generalized ARMA$(p_i,q_i)$-GARCH(1,1) process \cite{tsay}. This model is given by
\bea
r_{i,t} &\,=\,& \phi_0^i + \sum\limits_{j=1}^{p_i}  \phi^i_j \, r_{i,{t-j}} + a^i_t + \sum\limits_{k=1}^{q_i} \theta^i_k\, a_{i,{t-k}} \,, \notag \\
a_{i,t} &\,=\,& \sigma_{i,t} \, \epsilon_{i,t}  \,, \labell{6x3} \\
\sigma_{i,t}^2  &\,=\,& \alpha^i_0 + \alpha^i_1 \, a_{i,{t-1}}^2 + \beta^i_1 \, \sigma_{i,{t-1}}^2 \,. \notag
\eea
The first equation is the ARMA$(p_i,q_i)$ mean equation and other two equations are GARCH(1,1) volatility equations. $\phi^i_j$ is the auto-regressive coefficient and $\theta^i_j$ is moving average parameter. $\alpha^i_0$, $\alpha^i_1$ and $\beta^i_1$ are standard GARCH(1,1) parameters. We first find the `optimum' values of $p_i$ and $q_i$ by modeling the time series as ARMA process based on BIC measure \cite{forecast, autoarima1}. With these optimal orders, we simultaneously fit the model \eqref{6x3} and estimate all the free parameters \cite{rugarch}. As expected, we find that although there is a small difference in the numerical values, the main results in sections \ref{rmt}, \ref{eigenvalues} and \ref{eigenvectors} remain intact.

%%%%%%%%%%%%%%%%%%%%%%%%%%%%%%%%%%%
\section{Discussion} \labell{dis}

In this paper, we study the properties of eigenvalues and eigenvectors of volatility correlation matrix. We estimate volatility time series for 427 stocks in S\&P 500 by modeling the price fluctuations as GARCH(1,1) process. The empirical distribution for the estimated volatility fits well with the lognormal distribution. Using these volatility time series, we construct the volatility correlation matrix and study distribution of its eigenvalues. By fitting the empirical cumulative probability distribution in analytical expression, we find that approximately 40\% eigenvalues fall under the Mar\u{c}enko-Pastur distribution \eqref{eq2x1}. Whereas, for the same time period, approximately 75\% eigenvalues of return correlation matrix are within the analytical bounds from the RMT. We also find that the largest eigenvalue for the volatility correlation matrix is 237 whereas for return correlation matrix it is 193. This suggests that correlations in volatility of assets across the market are relatively stronger compared to correlations in the price fluctuations. To further establish that volatility eigenvalues falling under the Mar\u{c}enko-Pastur distribution are pure noise, we study the short and long-range correlations among eigenvalues in section \ref{eigenvalues}. Since volatility correlation matrix is a real symmetric matrix, we compare the statistical properties of eigenvalues with the analytical results from GOE. We find that for optimum values of unfolding parameters, the nearest-neighbor spacing distribution, next to nearest-neighbor spacing distribution and number variance fit well in theoretical results for GOE. These results strongly suggest that approximately 40\% eigenvalues of volatility correlation matrix carry no relevant information about the market.

In section \ref{evecdis}, we study the distribution of eigenvectors associated with eigenvalues within the RMT bounds. We find that eigenvector distribution is sharply peaked around zero as compared to a Gaussian distribution. We attribute this behavior to the strong correlations in volatility across the market. Once the market mode is removed from the volatility time series, we find that eigenvectors indeed have a distribution close to a Gaussian. In section \ref{ecosec}, we study the properties of eigenvectors which are outside the RMT distribution. These eigenvectors are supposed to carry genuine information about the market. Similar to return correlation matrix \cite{bouchaud1, stanley1}, we find that the largest eigenvector is the market mode. The largest eigenvectors of return correlation matrix are relatively stable in time and are dominated by only a few industry groups \cite{stanley5, stanley4}. We use inverse participation ratio of industry weight vectors \eqref{5x4} as an indicator of large contributions from a few industries and compare it for return and volatility eigenvectors. We find that in comparison with return eigenvectors, very few of largest volatility eigenvectors are dominated by a few industry group. This result is consistent with what is observed in practice. For instance, one can reduce risk of a portfolio by diversifying it across different economic sectors but it is much harder to remove volatility risk. Finally in section \ref{robust}, we discuss the robustness of our results. Since our results are based on estimation of $427\times 3$ parameters for all the GARCH(1,1) processes \eqref{eq1x3}, it is necessary to understand how they are affected when we have a very small or large number of parameters. Though there are small quantitative differences, our results remain invariant qualitatively in both scenarios.

In appendix \ref{volreturn}, we have also summarized results on application of RMT to correlations among volatility returns, which are defined as \eqref{cx1}. In this case, we found that the largest eigenvalue is 91 and 73\% eigenvalues are pure noise. Compared to return eigenvectors, there are less eigenvectors of volatility return correlation matrix which are dominated by a few industry groups. However, in comparison with volatility correlation matrix, these eigenvectors carry more information about the industry groups. In the five largest eigenvectors, we also find that dominating industry groups are same as what appear in the five largest return eigenvectors.

In this paper, we use one of the simplest methods, GARCH(1,1), to estimate daily volatility. However, there are various other proxies and methods of estimation. For example, there are other multivariate models of volatility \cite{engle02}. The implied volatility of an asset is estimated by observing the option prices in the market. One can also use high-frequency data to estimate variance or mean absolute deviation of prices. It will be interesting to apply tools from the RMT on all these indicators for a deeper understanding of correlations among volatility. Since implied volatility for an underlying asset varies both with the strike and maturity, we expect an even rich structure in correlations. This paper certainly provides a first step in these directions.

Note that the original applications of the RMT in financial market were on correlations among price fluctuations \cite{bouchaud1}. In this paper, we instead focus on the second moment of these fluctuations. A deeper understanding of financial market will require information about all the correlations among its components. Hence, it will be interesting to extend our work on realized higher moments, \eg skewness and kurtosis, which can be estimated empirically.

One of the main results of this paper is to show that a significant number of eigenvalues of volatility correlation matrix are pure noise. Since volatility correlation matrix has applications in risk management and forecasting \cite{implied, multivariate}, it will be natural to understand how volatility correlation matrix can be cleaned to carry only genuine information. The next step will be to see how this cleaned volatility correlation matrix improves the price forecast and risk estimates. We only list few here, but there is a vast literature where matrix cleaning methods are applied to financial correlations \cite{potters1, sharifi, daly1, daly2, bouchaud3, papp, urban}. It is expected that these tools will also find applications to volatility correlation matrix.

In section \ref{ecosec}, we compare the information contained in the largest eigenvectors about the GICS industry groups. We observe that very few of the largest eigenvectors of volatility correlation matrix are dominated by industry groups. Since we expect these eigenvectors to carry genuine information about the market, this result hints two possibilities. First, it has been observed for the eigenvectors of return correlation matrix that few largest eigenvectors are quite stable in time. Therefore, these eigenvectors are dominated by a particular industry group for a longer time period. As one moves to smaller eigenvalues, the time scale of stability begins to reduce. It is quite possible that for volatility correlation matrix, only two or three largest eigenvectors are stable and rest of them are just `random'. However, this possibility do not reconcile with the fact that compared to return correlations, volatility correlations are much stronger in the same time period. So this leads to a second scenario that volatility might organize itself in non-linear structures which are overlapped of different industry groups. Since these ideas  are useful in the volatility risk management and volatility arbitrage strategies, it will be very interesting to study time evolution and structure of volatility eigenvectors \cite{mantegna1, mantegna2, marsili1, coron1,stanley5, stanley4, allez1, allez2,fenn,conlon}.

\vskip .5cm

\noindent {\bf Acknowledgments:}
We thank Samuel Vazquez and John Nieminen for their insightful comments and suggestions. AS would like to thank Robert Myers for his support and encouragement in this work. AS also thanks Apurva Narayan and Heidar Moradi for several interesting discussions. This research was supported in part by Perimeter Institute for Theoretical Physics. Research at Perimeter Institute is supported by the Government of Canada through Industry Canada and by the Province of Ontario through the Ministry of Research and Innovation.

%APPENDIX
%%%%%%%%%%%%%%%%%%%%%%%%%%%%%%%%%%%%%%%%%%%%
\appendix

\section{Gaussian broadening and unfolded eigenvalues}
\labell{appa}

\begin{figure}
\centering
\subfigure[]{\includegraphics[width=.5\textwidth]{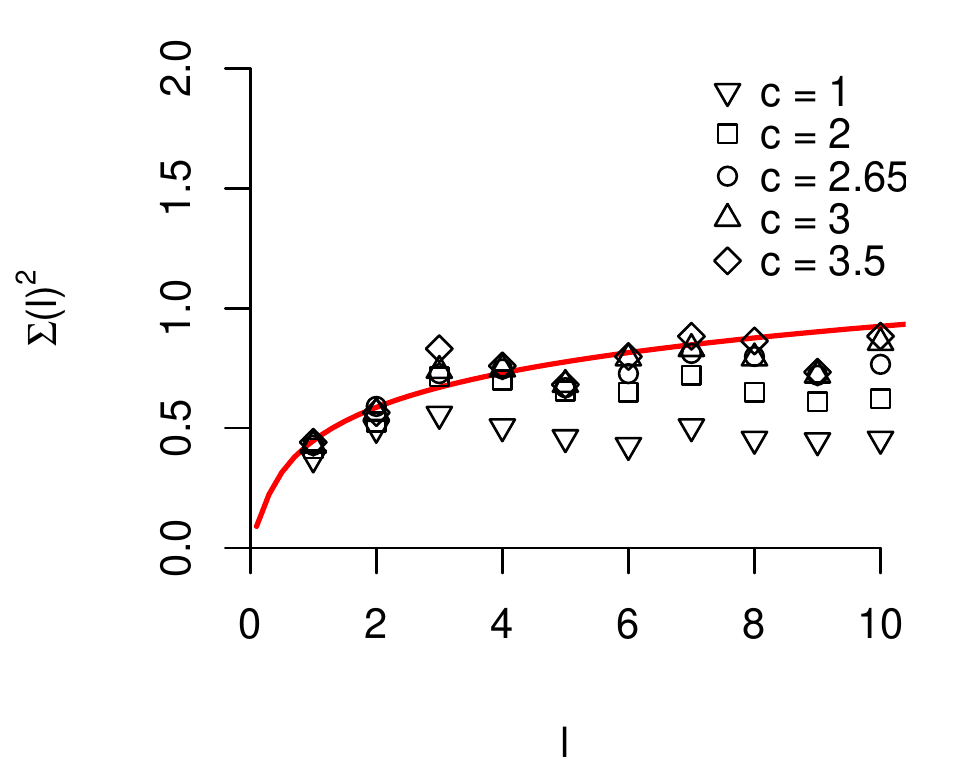}}
\caption{(Colour Online) We show number variance for different values of unfolding parameter $c$, keeping $w=0.0047$ fixed. The solid line is the theoretical value of number variance for GOE. We find that as we change $c$, the empirical estimates approach to the exact result for GOE. In a small range around the optimum value of $c=2.65$, empirical estimates are relatively stable. }
\label{fig10}
\end{figure}

In this appendix, we discuss the Gaussian broadening procedure for the eigenvalues of volatility correlation matrix. The empirical cumulative probability distribution for eigenvalues $\lambda_i$ of a random matrix is given by,
\be
F(\lambda_i) \,=\, {1\ov N} \sum \limits_{j=1}^N \Theta(\lambda_i-\lambda_j) \,.
\labell{ax1}
\ee
where $N$ is the total number of eigenvalues and $\Theta(\lambda_i)$ is the Heaviside step function. This probability distribution can be divided into two parts,
\be
F \,=\, F_{av}(\lambda) + F_{f}(\lambda)\,,
\labell{ax2}
\ee
where $F_{av}$ is the average part and $F_f$ is the fluctuating part, which is zero when averaged over the ensemble. Now one can get unfolded eigenvalues $\xi_i$ using the average part of the cumulative probability distribution,
\be
\xi_i \,=\, N F_{av}(\lambda_i)\,.
\labell{ax3}
\ee
The unfolded eigenvalues are a map from eigenvalues $\lambda_i$ to $\xi_i$, such that they have a uniform distribution. Also, note from \eqref{ax1} and \eqref{ax3}, $\xi_i$ are independent of $N$.

To separate $F_{av}$, from the estimated cumulative probability \eqref{ax1}, we use the procedure of Gaussian broadening \cite{broadening}. We replace delta function peaks at each eigenvalue $\lambda_i$ in \eqref{ax1} by a Gaussian distribution with mean $\lambda_i$ and standard deviation $\eta_i$. To estimate an optimum value of $\eta_i$ for each eigenvalue, we divide the eigenvalue scale in artificial `sub-bands' of width $w$ and number them as $m=\{1,2,\dots\}$. We calculate the average distance between eigenvalues in each sub-band and represent it by $d_m$. Then for each eigenvalue $\lambda_i$, we find the standard deviation
\be
\eta_i \,=\, 2\,c\, d_m \,,
\ee
where $c$ is a broadening parameter and $d_m$ is such that the eigenvalue $\lambda_i$ belongs to sub-band $m$.
%Note that the constant $c$ is a scale that tells on average to how many eigenvalues a particular eigenvalue will couple.

The values of sub-band width $w$ and broadening parameter $c$ are generally chosen to get the best fit for short-range and long-range correlations when compared with the theoretical results. For discussion in section \ref{eigenvalues}, where we compare the correlations among eigenvalues of volatility correlation matrix with GOE, we have used $w=0.0047$ and $c=2.65$. Note that in case of next to nearest-neighbor spacing distribution in section \ref{nnnbr}, we classify eigenvalues $\lambda_i$ in two different groups with even and odd $i$, and then unfold each group separately. So in these cases, to keep the correlation structure consistent with unfolding in sections \ref{nnbr} and \ref{number}, we use $\eta_i^{\textrm{\tiny even/odd}}=c\,d_m^{\textrm{\tiny even/odd}}$.

Finally, while comparing the properties of empirical data with RMT results, one needs to ensure that the agreement is not because of a particular choice of unfolding parameters. One should observe that as the unfolding parameters are varied in a particular range, the empirical results converge to theoretical values. Generally, short-range correlations are less sensitive to the unfolding procedure and this convergence is more visible in long-range correlations. In figure \ref{fig10}, we have shown number variance for various values of $c$, keeping $w=0.0047$ fixed. One can clearly see that as $c$ reaches the optimum value, the estimates for number variance approach to the theoretical values for GOE. We observe similar behavior when $w$ varies, keeping $c$ fixed.

%%%%%%%%%%%%%%%%%%%%%%
\section{Eigenvalues after removing the market mode} \labell{appb}

To remove the effect of the market mode, we regress the normalized volatility time series $\widehat{\sigma}_{i,t}$ with the market mode \eqref{5x1} as discussed in section \ref{evecdis}. As shown in \eqref{5x2}, the residual time series $\varepsilon_i$ can be obtained from
\be
\widehat{\sigma}_{i,t} \,=\, \alpha^i + \beta^i M_t +\varepsilon_{i,t}\,,
\labell{bx1}
\ee
which is further used to calculate the correlation matrix $\tilde{\mathbb{C}}$. The elements of correlation matrix $\tilde{\mathbb{C}}$ are related to original correlation matrix $\mathbb{C}$ as follows
\bea
\tilde{\mathbb{C}}_{ij}  & \,=\, & {\langle \varepsilon_i \varepsilon_j \rangle \ov \tilde{s}_i \tilde{s}_j} \\
& \,=\, & {\mathbb{C}_{ij} \ov \tilde{s}_i \tilde{s}_j} - {\beta_i \beta_j \lambda_N N \ov \tilde{s}_i \tilde{s}_j}\,.
\labell{bx2}
\eea
Here $\tilde{s}_i$ is the standard deviation for residual time series $\varepsilon_i$. In obtaining \eqref{bx2}, we have also used $\langle \varepsilon_i \rangle=0$, $\langle \varepsilon_i M\rangle=0$ and that variance of time series $\widehat{\sigma}_{i,t} $ is one. If market is weakly correlated, $\lambda_N$ will not be very large. Also, the dynamics of overall market will have less influence on individual time series and coefficients $\beta_i$ will be small. In this case, if $\tilde{s}_i=\tilde{s}$ for all $i$, to the leading order from \eqref{bx2},
\be
\tilde{\lambda}_i \,=\, {\lambda_i \ov \tilde{s}^2}\,.
\labell{bx3}
\ee
Also, let us assume that only the market mode is outside the the Mar\u{c}enko-Pastur distribution. In this case, all the $\beta_i$ are approximately equal and $\beta_i \approx 1/N$. Using this in the variance of residual time series $\varepsilon_i$,
\be
\tilde{s}^2 \approx 1 - {\lambda_N \ov N} \,.
\ee
This can further be used in \eqref{bx3} to get $\tilde{\lambda}_i \approx {\lambda}_i N/(N-\lambda_N)$. Note that this is consistent with the arguments from the normalization of eigenvalues in the discussion after equation \eqref{5x2}. However, if $\lambda_N$ is very large, the corrections to \eqref{bx3} will not be negligible. In case of volatility correlation matrix, the largest eigenvalue is of the order of $N$. Hence, although $\tilde{\lambda}_i$ are close to $\lambda_i N/(N-\lambda_N)$, we observe that there are noticeable corrections.

%%%%%%%%%%%%%%%%%%%%%%%%%%%%%%%%%%%%%%%%%%%
\section{Correlations among volatility return} \labell{volreturn}

\begin{figure}
\centering
\subfigure[]{\includegraphics[width=.5\textwidth]{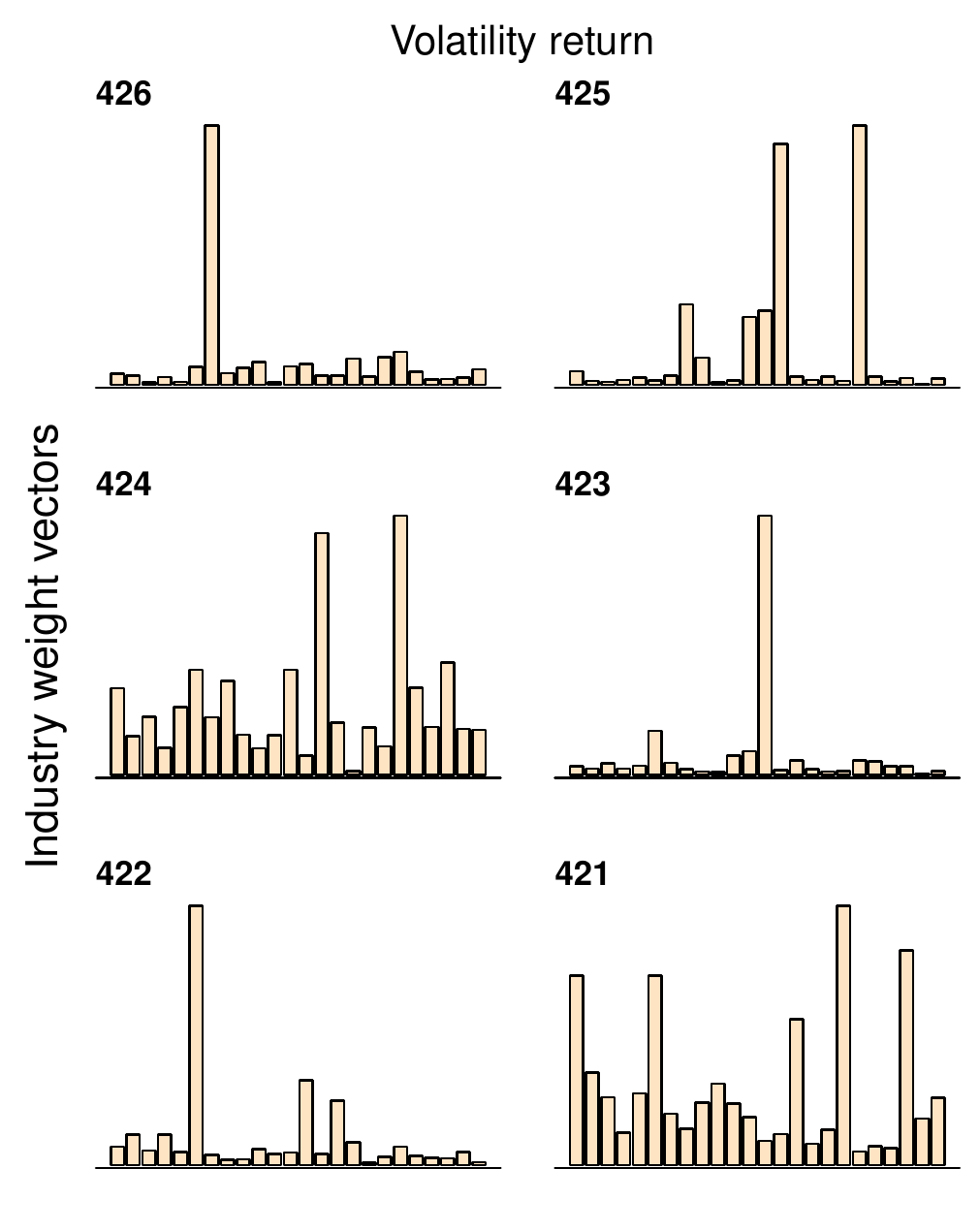}}
\caption{(Colour Online) We have shown the components of the weight vectors corresponding to six largest eigenvectors of volatility return correlation matrix. For volatility return eigenvectors, the largest contributions come from following industry groups: 426 -- utilities, 425 -- banks and energy, 424 -- banks and energy, 423 -- real estate, 422 -- semiconductors and semiconductors equipments, 421 -- Household \& Personal Products. Note that the industry groups in five largest eigenvectors match with the dominating groups in return eigenvectors, that are shown in figure \ref{fig7}.}
\label{fig8}
\end{figure}

\begin{figure}
\centering
\subfigure[]{\includegraphics[width=.5\textwidth]{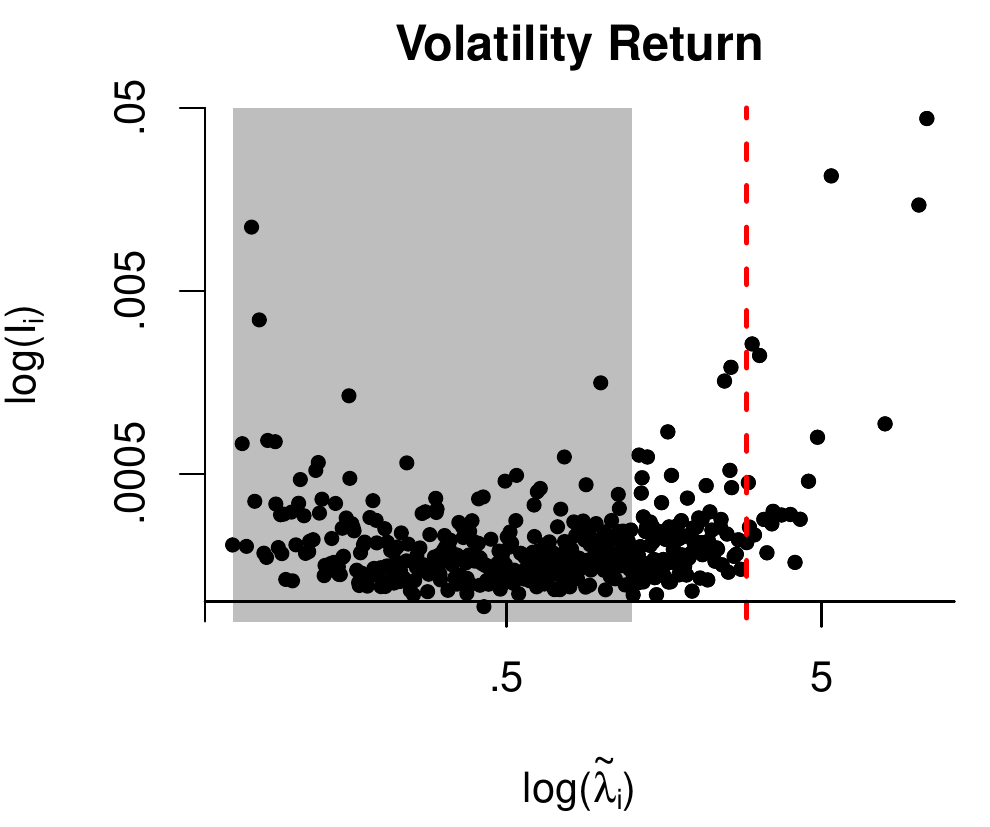}}
\caption{(Colour Online) We have plotted inverse participation ratio of weight vectors for volatility return eigenvectors. On the log-log plot, the x-axis is eigenvalues that we get after removing the influence of the market. The vertical dashed line shows the location beyond which, the twenty largest eigenvalues fall. We can see that compared to return correlation matrix, less eigenvalues have large inverse participation ratio. However, compared to volatility correlation matrix, more eigenvectors receive dominating contributions from industry groups.}
\label{fig9}
\end{figure}

In this appendix, we briefly mention our results on the application of RMT on volatility return. Using volatility time series $\sigma_{i,t}$ in \eqref{eq1x3}, we can define the volatility return
\be
\delta \sigma_{i,t} \,=\, \log \left( {\sigma_{i,t+1} \ov \sigma_{i,t}} \right) \,.
\labell{cx1}
\ee
These volatility return time series are further normalized such that they have zero mean and unit variance. We can arrange these time series in $N \times (T-1)$ matrix $\mathcal{G}$ and define the correlation matrix
\be
\mathcal{C} \,=\, {1\ov T-1} \mathcal{G}\mathcal{G}^T\,.
\labell{cx2}
\ee
For this matrix, the smallest eigenvalue is $.05$ and largest eigenvalue is $91$. We can follow the procedure in section \ref{rmt} and find that now 73\% eigenvalues fall under the Mar\u{c}enko-Pastur distribution. The statistical properties of eigenvalues are also consistent with that of GOE.

We further find that eigenvectors corresponding to eigenvalues within the RMT bounds have a Gaussian distribution. The largest eigenvector is again the market mode. Following procedure in section \ref{evecdis}, we can remove the influence of the market mode and using projection matrix \eqref{5x3}, we can calculate the industry weight vectors. The measure \eqref{5x7} for the merit of linear model \eqref{5x2} for the volatility return correlations is $K_{vr}=0.32$. In figure \ref{fig8}, we have shown weight vectors for six largest eigenvectors, excluding the market mode. Once again, in comparison to return eigenvectors, there are less volatility return eigenvectors which are dominated by a few industries. For the weight vectors, we can also calculate inverse participation ratio and it is shown in figure \ref{fig9}. We can further compare the inverse participation ratios with the benchmark value $I_0$, that we defined in section \ref{ecosec}. We find that twelve among the twenty largest eigenvectors of volatility return correlation matrix have an inverse participation ratio less than this benchmark value. This number is less than sixteen, that we get for volatility correlation matrix. This indicates that for the data we have discussed, volatility return eigenvectors contain more information about the industry groups as compared to volatility eigenvectors. We also notice that the dominating industry groups in five largest eigenvectors for return and volatility return correlation matrices are the same.

%%%%%%%%%%%%%%%%%%%%%%%%%%%%%%%%%%%%%%%%%%%%
\bibliographystyle{unsrt}
\bibliography{references}{}

\end{document}